\documentclass[12pt]{article}
\usepackage{amsmath, amssymb}

\usepackage{epsfig}

\newtheorem{theorem}{Theorem}[section]
\newtheorem{proposition}[theorem]{Proposition}
\newtheorem{corollary}[theorem]{Corollary}
\newtheorem{lemma}[theorem]{Lemma}

\newcommand{\rd}{{\rm d}}
\newcommand{\be}{\begin{equation}}
\newcommand{\ee}{\end{equation}}
\newcommand{\bey}{\begin{eqnarray}}
\newcommand{\eey}{\end{eqnarray}}

\newcommand{\tri}{| \! |\!|}

\newcommand{\eqn}{\begin{eqnarray}}
\newcommand{\eeqn}{\end{eqnarray}}

\newcommand{\Tor}{{\bf T}}

\renewcommand{\a}{\alpha}

\newcommand{\e}{\varepsilon}

\newcommand{\om}{{\omega}}

\newcommand{\cU}{{\cal U}}

\newcommand{\bR}{{\mathbb R}}

\newcommand{\bN}{{\bf N}}
\newcommand{\bZ}{{\mathbb Z}}

\newcommand{\wt}{\widetilde}
\newcommand{\wh}{\widehat}

\newcommand{\cI}{{\cal I}}

\newcommand{\cG}{{\cal G}}

\newcommand{\cD}{{\cal D}}

\newcommand{\cN}{{\cal N}}

\newcommand{\cT}{{\cal T}}

\newcommand{\ZK}{\cG}

\date{Apr 6, 2006}

\begin{document}

\title{Decay of the Fourier transform of surfaces with
\\
vanishing curvature}
\author{L\'aszl\'o Erd\H os${}^1$\thanks{Partially
supported by  EU-IHP Network ``Analysis
and Quantum'' HPRN-CT-2002-0027}
\\ Manfred  Salmhofer${}^2$\thanks{Partially supported by DFG grant
Sa 1362/1--1 and an ESI senior research fellowship.}
\\
\small
${}^1\;$Institute of Mathematics, University of Munich, 
\\
\small
Theresienstr. 39, D-80333 Munich, Germany
\\
\small
${}^2\;$Max--Planck Institute for Mathematics, Inselstr.\  22, 04103 Leipzig,
and
\\ 
\small
Theoretical Physics, University of Leipzig, Postfach 100920, 04009 Leipzig, 
Germany
\\
}

\maketitle

\abstract{We prove $L^p$-bounds on the Fourier
transform of measures $\mu$ supported on two dimensional
surfaces. Our method allows to consider surfaces
whose Gauss curvature vanishes on
a one-dimensional submanifold.  Under a certain
non-degeneracy condition, we prove that $\wh\mu\in L^{4+\beta}$, $\beta>0$,
and we give a logarithmically divergent bound on the $L^4$-norm.
  We use this latter bound to estimate almost singular integrals
involving the dispersion
relation, $e(p)= \sum_1^3 [1-\cos p_j]$,
of the discrete Laplace operator on the cubic lattice.
We briefly explain our motivation for
this bound originating in the theory of random Schr\"odinger
operators.
}


%\tableofcontents

\section{Introduction}

\subsection{Notations and background}

Let $\Sigma$ be a smooth, compact hypersurface 
embedded in $\bR^3$ or in the torus $\Tor^3= [-\pi, \pi]^3$.
 Let $\rd m$ be the induced surface area
 measure on $\Sigma$ and let $f\in C_0^\infty(\Sigma)$, i.e.\ $f$ is 
a smooth function supported away from the boundary of $\Sigma$. 
%compact support.
 Let $\kappa_i=\kappa_i(p)$, $i=1,2$, denote
the two principal curvatures at $p\in \Sigma$,  let
$K= \kappa_1\kappa_2$ be the Gauss curvature and  $H=\kappa_1+\kappa_2$
the mean curvature.

 We define the Fourier transform of
the measure $\rd\mu = f \rd m$,
\be
      \wh\mu(\xi)= \int_\Sigma e^{i\xi\cdot p} f(p) \rd m(p) \; , \qquad 
    \xi\in \bR^3,
\label{Jdef}
\ee
and we investigate the decay properties of $\wh\mu$ at infinity.
We  prove that 
\be
     \wh\mu\in L^{4+\beta}(\bR^3), \qquad \beta>0\; ,
\label{4-e}
\ee
and
\be
    J_\eta=
 \int_{|\xi|\leq \eta^{-1}} |\wh\mu(\xi)|^4 \rd \xi \leq C|\log\eta|^{10},
\label{finalgoal}
\ee
with some constant $C$ depending only on $f$ and on a few geometric
properties of the surface $\Sigma$.
The estimate \eqref{finalgoal} indicates a decay 
$\wh\mu(\xi) \lesssim \langle \xi\rangle^{-3/4}$, with
$\langle \xi \rangle  = (\xi^2+1)^{1/2}$, for almost all $\xi$.

By a standard stationary phase argument 
(see, e.g., Theorem 1, Section VIII.3.1 of \cite{St}) it is well known
that the decay estimate
\be
  |\wh\mu(\xi)|\leq \frac{C}{\langle \xi\rangle^r}
\label{triv}
\ee 
holds 
with $r=1$ 
if  $K$ nowhere vanishes  on 
the support of $ f$.
The constant depends on the lower bound on $|K|$ and on
supremum bounds of a few derivatives of $ f$. In particular,
this bound holds for uniformly convex surfaces.

Fewer results are available if $K$ is allowed to vanish.  
In the
extreme case, when $K\equiv 0$ and $\Sigma$ is flat, 
$\wh\mu(\xi)$ does not decay  in the direction orthogonal to $\Sigma$.
In the general case,
local results can easily be obtained by a stationary phase
analysis.
To formulate them, let $\nu(p)$
denote the  unit normal at the point $p\in \Sigma$.
Assume that $f$ is supported in a sufficiently small neighborhood $\cU$
of $p$. 
Then the rate of decay of $|\wh\mu (\lambda \nu(p))|$ for $|\lambda| \gg 1$
is  estimated as
\be
    |\wh\mu (\lambda \nu(p))| \le C|\lambda|^{-k/2}, \qquad |\lambda|\gg 1,
\label{kcurv}
\ee
where $k$ is the number of nonvanishing principal curvatures at $p$
(see e.g. Section VIII.5.8 of \cite{St}).
The constant in this estimate
depends on the point $p$ unless
a uniform lower bound is known on the non-vanishing  curvatures. For example,
\eqref{triv} holds with a uniform constant  and with $r=1/2$
if $|\kappa_1|+|\kappa_2|\ge c_0>0$ on the support of
$ f$. For vectors $\xi$ that are not parallel with any normal
vector $\nu(p)$, $p\in \cU$,
 the decay rate is polynomial with arbitrary high degree,
\be
     |\wh\mu(\xi)| \le C  |\xi|^{-N}, \qquad N\in \bN \; .
\label{Ncurv}
\ee
Here the constant depends on $N$ and on  $\inf\{ |\nu(p)\times\om_\xi  |
\; : \; p\in \cU\}$ with $\om_\xi= \xi/|\xi|$, where
 $\times$ denotes the cross product.

To obtain an $L^p$-bound on $\wh\mu$, one must control
the  behavior of the constants in \eqref{kcurv} and \eqref{Ncurv}
that depend on further geometric
properties of $\Sigma$. 
For a convex hypersurface $\Sigma$, 
Bruna, Nagel and Wainger \cite{BNW} have shown
that 
$$
|\wh\mu (\lambda \nu(p))| \leq C \; \mbox{Vol}( \, B(p, \lambda^{-1}) \, ) \; ,
$$
where $B(p, h)= \{ y\in \Sigma \; : \; (p-y)\cdot \nu(p) \leq h\}$
is the spherical ``cap'' of height $h$ ($h\ll 1$) around $p$
and $\nu(p)$ is the ``outer'' normal.
One can thus determine the constant in \eqref{kcurv}
 by a local Taylor expansion.
Iosevich \cite{Ios} showed that $|B(p, \delta)|\leq C\delta^r$
is equivalent to  \eqref{triv} for convex hypersurfaces of finite type
(i.e. the order of contact with any tangent line is finite).
The convexity is essential in these estimates.

\medskip
\noindent

Our goal is to prove \eqref{4-e} and
\eqref{finalgoal} for a class of non-convex
hypersurfaces; in particular  $K$ will be allowed to vanish on a 
one-dimensional submanifold of $\Sigma$. We assume
that both curvatures cannot vanish at any point, i.e. there
is no flat umbilic point on $\Sigma$. By compactness this means 
\be
|\kappa_1|+|\kappa_2| \ge (const) >0\; ,
\label{sumk}
\ee
in particular $|\wh\mu (\xi)|\leq C |\xi|^{-1/2}$ for all
$\xi$, and $|\wh\mu(\xi)|\le C(\om_\xi) |\xi|^{-1}$
(with an $\om_\xi$-dependent constant)
 unless $\xi$
is parallel with a normal vector $\nu(p)$ on the zero set
of $K$, i.e. $K(p)=0$. 
If one naively uses the estimate
$|\wh\mu (\xi)|\leq C |\xi|^{-1/2}$ for all  $\xi$'s on
 the two-dimensional
submanifold $\{ \xi \; : \; \xi \Vert \nu(p),\;  K(p)=0\}\subset \bR^3$
and $|\wh\mu(\xi)|\le C(\om_\xi) |\xi|^{-1}$ for all other $\xi$, 
then the integral
$J_\eta$ diverges as $|\log\eta|$. This indicates that
the bound \eqref{finalgoal} is close to optimal
for surfaces satisfying \eqref{sumk}.
For the proof, however,  we will need further technical non-degeneracy
 assumptions on 
$\Sigma$.

Note that this argument is only heuristic since it neglects  to control
the constant in $|\wh\mu(\xi)|\le C(\om_\xi)|\xi|^{-1}$.
 The main technical result (Theorem \ref{prop:noc})
is to give an effective
estimate for $|\wh\mu(\xi)|$ 
that can be integrated to obtain \eqref{4-e}, \eqref{finalgoal}
(Corollary \eqref{cor:L4}).

\bigskip
\noindent
We mention that  the lack of decay due to the vanishing
curvature can be mitigated by a curvature factor
in the integral. The following general
result was obtained by Sogge and Stein \cite{SS}
$$
    \Big| \int_\Sigma e^{i\xi\cdot p} K(p)^4   f(p) \rd m(p) \Big|
 \leq C \langle \xi \rangle^{-1}
$$
for any hypersurface.
Similar result holds for hypersurfaces in any dimension.

We also mention that the bound \eqref{triv}
with some $r>0$ 
implies  classical Fourier restriction estimates, for example
$$
   \Big( \int_\Sigma |\wh g|^2\rd \mu\Big)^{1/2}
  \leq C \|g\|_{2(r+1)/(r+2)} \; 
$$
for any function $g$ on $\bR^3$ \cite{Gr}.
The restriction theorem 
has been investigated for 
certain special surfaces with vanishing curvature.
Oberlin considers a rotationally symmetric surface
with curvature vanishing at one point \cite{Ob}.
Very recently Morii obtained a restriction theorem
for surfaces given as graphs of real polynomials
that are sums of monomials \cite{Mo}. 
It would be interesting to investigate 
the restriction
theorem  for the class of hypersurfaces we consider.

\bigskip
\noindent

The paper is organized as follows. In Section
\ref{sec:motiv} we explain our original motivation
to study this problem.
In Section \ref{sec:gense} we formulate the 
 assumptions on the surface $\Sigma$ and state our bound on the
decay of $|\wh\mu(\xi)|$. As a corollary
of this estimate, we will obtain \eqref{4-e} and \eqref{finalgoal}.
In Section \ref{sec:4denom} we formulate a theorem,
the so-called Four Denominator Estimate, that is 
ultimately connected with the $L^4$-bound
of the Fourier transform of an explicitly given surface.
This surface is the level set of the dispersion
relation of the discrete Laplace operator (see \eqref{edef} below).
Section \ref{sec:proofthm} contains the proof of the
bound on  $|\wh\mu(\xi)|$.
Finally, in Section \ref{sec:fourden} we prove the
Four Denominator Estimate. It will be an easy
consequence of our general bound on $|\wh\mu(\xi)|$,
once we have checked that the assumptions are satisfied
for  this particular surface. Despite the explicit
formula for the dispersion relation, verifying
the otherwise  generic assumptions is 
a non-trivial task.

\subsection{Motivation: Random Schr\"odinger evolution}
\label{sec:motiv}

Although the decay of the Fourier transform of measures
 supported on hypersurfaces  is an interesting and
 broadly studied problem itself,
our motivation to prove the estimate \eqref{finalgoal}
came from elsewhere.

% We explain briefly and informally
%the origin of our interest in this question.

%\subsubsection{Random Schr\"odinger operators}

 We studied the long-time behavior
of the random Schr\"odinger equation
\be
   i\partial_t \psi_t(x) = \Big[ -\frac{1}{2} \Delta_x + \lambda V(x)
 \Big] \psi_t(x) \; ,   \qquad \psi_t(x)\in L^2(\bR^3),
\label{Schr}
\ee
in the three dimensional Euclidean space, $x\in \bR^3$.
Here $V(x)$ is a random potential with a short scale
correlation and $\lambda$ is 
a small coupling constant. 

The equation \eqref{Schr} models the quantum evolution
of an electron in a random impure environment.
It has been proved that the electron is localized
for sufficiently large $\lambda$.
It is  conjectured, but not yet proven, that the evolution is delocalized,
moreover diffusive for all times,
 if $\lambda$ is sufficiently small. In \cite{ESYI}, \cite{ESYII},
jointly with H.-T. Yau
we proved a weaker statement, namely we 
proved diffusion up to time scale $t\sim \lambda^{-2-\kappa}$, $\kappa>0$,
in the scaling limit $\lambda\to 0$.
For the precise statement, the physical background and references, see
\cite{ESYI}.

The discrete analogue of \eqref{Schr} is the celebrated Anderson
model \cite{A}. In this model the electron is hopping on the lattice,
 $x\in \bZ^3$, generated by the
discrete Laplace operator $\Delta_x$. The random potential
$V(x)$ describes the potential strength  of a random obstacle
at the location $x$. It is given by a collection of
i.i.d. random variables $\{ V(x)\; : \; x\in \bZ^3\}$.
 Since the (de)localization
problem concerns large distances,
physically there is no difference
between the continuous and the discrete model. In fact, 
the proofs in the localization regime have technically been
somewhat simpler for the discrete model since the large 
momentum regime is not present.  
Similar simplifications have  arisen when we implemented
 our diffusion result \cite{ESYI}, \cite{ESYII}
to the discrete setup \cite{ESYdisc}. However,
the lattice formulation gave rise to a
seemingly innocent technical
difficulty that became an unexpectedly  tough problem.

\medskip
\noindent
The basic approach of our work on random Schr\"odinger evolutions
 is perturbative:
we expand the unitary kernel, $e^{-itH}$, 
 of $H=-\frac{1}{2} \Delta + \lambda V$
around the free evolution, $e^{it\Delta/2}$.
After taking the expectation with respect to
the randomness, the Wigner transform of $\psi_t$
is written as a sum over Feynman graphs representing
different collision histories. The value of each 
Feynman graph is a  multiple integral
of momentum variables $p_j\in \Tor^3$
that are subject to linear constraints. The integrand is a  product
of functions of the form $(\alpha - e(p_j) + i\eta)^{-1}$,
the so-called {\it time-independent free propagators}.
Here $\alpha\in \bR$ and the function $e(p)$ is 
the Fourier multiplier of $-\frac{1}{2}\Delta$.
The regularization $\eta$ is the inverse time, $\eta= t^{-1} \ll 1$.

One of our key steps is to prove that the evolution
becomes Markovian as $\lambda\to 0$. If the electron
collides with the same random obstacle more than once,
then Markovity is violated. We must thus  prove
that the  Feynman graphs with
 recollision processes have  negligible contributions. 
In the Feynman integral, a double recollision 
corresponds to a factor
\be
      \frac{\delta(p-q+r-v) \; \rd p \rd q\rd r \rd v}{(\a -e(p)+i\eta)
(\a -e(q)+i\eta)
(\a -e(r)+i\eta)(\a -e(v)+i\eta)}\; ,
\label{factor}
\ee
where $p,q$ and $r,v$ are the pre- and postcollision
velocities in the first and second
collisions with the same obstacle. The delta function
expresses a natural momentum conservation
(for more details on Feynman graphs,
see  \cite{ESYdisc}).
It is therefore necessary to give a
good estimate for the integral of these four denominators
connected with a delta function.
This will be our Four Denominator Estimate formulated
in Theorem \ref{thm:4de}.

\medskip

{\it Acknowledgement.} This work is  part
of a joint project with H.-T. Yau on quantum diffusion.
The authors express their gratitude for his discussions 
and comments on this work .
The authors are also indebted to A. Sz{\H u}cs
for helpful discussions.

\section{Statement of the main results}
\label{sec:geom}
\setcounter{equation}{0}

\subsection{Theorems on the decay of the Fourier transform.}\label{sec:gense}

In this section we formulate the geometric assumptions
and we state  a general theorem on the decay of
the Fourier transform of measures supported on surfaces.
First we discuss the case of a family of surfaces
that is represented as level sets of a regular function.
Then we explain how this result can be used
to investigate the case of a single surface.

Let $e(p)$ be a smooth real function on $\bR^3$ or $\Tor^3=[-\pi,\pi]^3$
 and let
$\Sigma_a =\{ p \; : \; e(p)=a\}$ be the $a$-level set
for any $a\in \bR$. Let $\cI\subset \bR$ be a finite union of 
compact intervals
such that the preimage $\cD=e^{-1}(\cI)$ is compact and
$\Sigma_a$ is a two-dimensional submanifold for each $a\in \cI$.
Let $f$ be a smooth function on $\cD$, 
and define
\be
    \wh\mu_a(\xi)= \int_{\Sigma_a} e^{i\xi\cdot p}
   f(p) \rd m_a(p)\; ,
\label{whmu}
\ee
the Fourier transform of the measure $ f \rd m_a$,
where $\rd m_a$ is the induced surface area measure
on $\Sigma_a$. 

We define
\be
   C_0= \mbox{diam}(\cD), \qquad   C_1= \| e \|_{C^5(\cD)} \; .
\label{C1def}
\ee

We set the following
\be
\mbox{\bf Assumption 1:}\qquad     C_2=\min_{\cD} | \nabla e| >0 \; ,
\qquad\qquad\qquad
\label{ass1}
\ee
i.e. we require that the level surfaces $\Sigma_a$, $a\in \cI$,
form a regular foliation of $\cD$. In particular, $\Sigma_a$ has no
boundary.

Let $K:\cD\to \bR$ be the Gauss curvature of
the foliation, i.e. $K(p)$ is the Gauss curvature
of $\Sigma_a$ at $p$ if $p\in \Sigma_a$.
Since $K$ is the determinant of the second
fundamental form of a smooth foliation, it
is a smooth function on $\cD$.

%We assume that $\nabla K$ does not vanish on the 
%set $\Gamma = \{  p \in \cD \; : \; K(p)= 0\}\subset \cD$, 
%so that in particular the zero set of $K$
%is  a two-dimensional submanifold in $\cD$.
%We actually make a stronger assumption requiring that
%the surface $\Gamma $  intersects
%the foliation $(\Sigma_a)_{a \in \cI}$ transversally:

We also assume that the zero set of the Gauss curvature  
intersects the foliation $(\Sigma_a)_{a \in \cI}$ transversally:

\medskip\noindent
{\bf Assumption 2. } Let $\ZK = \{  p \in \cD \; : \; K(p)= 0\}$.
Then 
\be
%\mbox{\bf Assumption 2:} \qquad
C_3=\min \big\{ |\nabla e(p) \times \nabla K(p)|
  \; : \;  p\in \ZK \big\} >0 \; .
\label{ass2}
\ee
This implies in particular that $\nabla K$ does not vanish on $\ZK$, 
so $\ZK $ is  a two-dimensional submanifold of $\cD$, 
and 
the two principal curvatures $\kappa_1(p)$, $\kappa_2(p)$
cannot vanish simultaneously, i.e.
there is no flat umbilic point. By compactness,
\be
\underline{\kappa} = 
   \min_\cD  \; ( |\kappa_1| + |\kappa_2| )  >0 \; 
\label{kappasum}
\ee
and %the lower bound 
$\underline{\kappa} $ depends only on $C_0, C_1, C_2, C_3$.
%This also guarantees that the principal curvatures %, $\kappa_1(p)$, $\kappa_2(p)$
%and the principal curvature directions 
% depend smoothly on $p\in\cD$ with uniform bounds on the derivatives.
Since $e(p)$ and $K(p)$ are smooth,  it follows from \eqref{ass2}
 that the zero curvature set, 
$$
   \Gamma_a  = \ZK \cap \Sigma_a %\{ p\in \Sigma_a \; : \; K(p) =0\},
$$
is a finite union of disjoint regular curves on $\Sigma_a$
for each  $a\in\cI$. All these curves are simple and closed.
Let
\be
p \mapsto
w(p) =  \frac{\nabla e(p) 
\times \nabla K(p)}{|\nabla e(p) \times \nabla K(p)|} \; 
\label{wdef}
\ee
be the unit vectorfield  tangent to $\Gamma_a$.

Define the normal map $\nu:\cD \to S^2$, given by
\be
   \nu(p)= \frac{\nabla e(p)}{|\nabla e(p)|} \; .
\label{def:nor}
\ee
The Jacobian of the normal map restricted to each surface,
$\nu:\Sigma_a\to S^2$, is the Gauss curvature,
$\mbox{det} \; \nu'(p)=K(p)$. 
%In particular, the normal map
%is a local bijection on each $\Sigma_a$ away from $\Gamma_a$
%with an inverse function $\nu^{-1}$, whose Jacobian matrix satisfies
%\be
%    \| (\nu^{-1})'(p)\| \leq \frac{C}{|K(p)|} \;, \qquad p\in 
%\Sigma_a\setminus\Gamma_a \; .
%\label{invjac}
%\ee
%The constant depends on $C_0,\ldots C_3$.

\medskip\noindent
{\bf Assumption 3. }
The number of preimages of $\nu:\Sigma_a\to S^2$ is finite, i.e.
\be
    C_4= \sup_{a\in \cI}
\sup_{\om\in S^2} \mbox{card} \{ p\in \Sigma_a\; :
 \; \nu(p) =\om\}  <\infty\; .
\label{ass3}
\ee

\bigskip\noindent
On the (union of) curves $\Gamma_a$, exactly one of the principal curvatures 
vanish, hence the principal direction  of the zero curvature
is well defined. This defines a (local) unit vectorfield
 $Z\in T\Sigma_a$ along $\Gamma_a$ in the tangent plane of $\Sigma_a$.
$Z$ is actually defined in a neighbourhood of $\Gamma_a$ as the direction
of the principal curvature that is small and vanishes on $\Gamma_a$. 
The orientation of $Z$ plays no role.
We assume that $Z$ is transversal to $\Gamma_a$ apart
from finitely many points (called {\it tangential points})
 and the angle between $Z$ and $\Gamma_a$ 
increases linearly near these points:

\medskip\noindent
{\bf Assumption 4:}
{\it 
There exist positive constants $C_5, C_6$ such that
for any $a\in \cI$  the set of tangential points, 
$$
    \cT_a= \{ p\in \Gamma_a \; : \; Z(p) \times w(p) = 0\} ,
$$ 
is finite with cardinality 
$ N_a = |\cT_a|\leq C_5$.
 For all  $p\in\Gamma_a$ 
\be
   |Z(p) \times w(p)| \ge 
 C_6 \cdot d_a(p) \; ,
\label{ass4}
\ee
where $d_a(p)$ is defined as follows. If $N_a=0$, then
 $d_a(p)=1$. If $N_a\neq 0$,
and
 $\cT_a =\{ p^{(1)}_a, p^{(2)}_a, \ldots , p^{(N_a)}_a\}$,
then  
\be
d_a(p)=\min\{ |p-p^{(j)}_a|\; : \; j=1, 2,\ldots N_a
    \} \; , \quad
  a\in \cI, \;\; p\in \Sigma_a \; .
\label{dadef}
\ee
}

\medskip\noindent
Alternatively, Assumption 4 can also be formulated by using 
the Hessian matrix $e''(p)$ of the function $e$.
At every point $p\in \Sigma_a$, $a\in \cI$,  we define the projection
$P=P(p)= I-|\nu\rangle\langle \nu|$ in the
three dimensional tangent space
 $T_p \bR^3$ onto the subspace orthogonal to the normal
vector $\nu=\nu(p)$.
The first order
variation of the normal vector at $p$ is
$$
    \nu(p+\rd p)-\nu(p) = |\nabla e(p)|^{-1}\; Pe''(p)P \; 
\rd p + O(\rd p^2)\; , \quad p\in \Sigma_a, \;\; \rd p \in T_p\Sigma_a\; ,
$$
i.e. $Pe''(p)P$ is proportional to the derivative of 
the Gauss map. It is easy to see that
Assumption 4 is equivalent to
\be
\mbox{\bf Assumption 4*:} \qquad\qquad
\forall p\in \Gamma_a \;: \; \qquad
    \| Pe''(p)P \, w(p)\|\ge C_6'\cdot d_a(p)\; 
\quad 
\label{ass4*}
\ee

\medskip\noindent
In the sequel, we work under the Assumptions 1--4.
We will use the notation $C^*$ and $c^*$ for various
large and small positive constants  that depend on
 $C_0, C_1, \ldots, C_6$ and $f$
and whose value may differ from line to line.
We define
\be
   D_a(\om)= \min \Big\{ |\nu(p_a^{(j)})
 \times\om| \; : \; 1\leq j\leq N_a
\Big\} \; , \quad  \om\in S^2\; 
\label{Ddef}
\ee
if $N_a\neq 0$ and $D_a(\om)=1$ if $N_a=0$.
The main technical result is the following 

\begin{theorem}\label{prop:noc} 
Under the Assumptions 1--4,  
there is $C^* > 0$ such that for all $a \in \cI$ and all $r>0$, $L \ge 1$ and $\om\in S^2$,
\be
    |\wh\mu_a(r\om)|\leq  C^*\Bigg\{  2^{-L} +
   \frac{1}{\langle r\rangle}+
   \frac{L^2}{  \langle r^{3/4}|D_a(\om)|^{1/2}\rangle}\Bigg\} \; ,
\label{fourdec}
\ee
and for any $0<\beta<\frac{1}{2}$
\be
    |\wh\mu_a(r\om)|\leq  C^*\Bigg\{ 
   \frac{1}{\langle r\rangle}+
   \frac{\beta^{-2}}{  \big\langle r^{\frac{3}{4}-\beta}
 |D_a(\om)|^{\frac{1}{2}-\beta}\big\rangle}\Bigg\} \; .
\label{fourdecbeta}
\ee
The positive constant  $C^*$ 
is uniform in $a\in\cI$. It depends  on the constants  
$C_0,  \ldots,  C_6$ and on the $C^2$ norm of $f$ in $\cD$.
\end{theorem}

\begin{corollary}\label{cor:L4}
Suppose that Assumptions 1--4 hold.
Then for any $M\ge 2$ 
\be
\sup_{a\in \cI} \int_{|\xi|\leq M} |\wh\mu_a(\xi)|^4
 \rd \xi \leq C^*|\log M|^{10}\;  .
\label{l4}
\ee
Moreover, for any $\beta>0$ we have 
\be
     \sup_{a\in \cI}
\| \wh\mu_a\|_{L^{4+\beta}(\bR^3)} \leq C^* \langle \beta^{-\frac{5}{2}}\rangle
\label{4b}
\ee
for the $L^{4+\beta}$ norm of $\wh\mu_a$.
\end{corollary}
 
\noindent
We have formulated our theorem for a family of
level surfaces $\Sigma_a=\{ p\; : \; e(p)=a\}$ of a given smooth
function $e(p)$
since we need the Four Denominator Estimate uniformly in $a$.
Our proof, however,  can  directly be applied to the decay
of the Fourier transform of a measure $\rd\mu=f\rd m$ on
a single smooth and compact surface $\Sigma$ in $\bR^3$.
We can allow $\Sigma$ to have a non-trivial boundary.
We formulate the necessary modifications and leave
the proof to the reader.

Let
 $\nu(p)$ be the unit normal vector at $p\in \Sigma$,
$\Gamma= \{ p\in \Sigma \; : \; K(p)=0\}$ be the
zero set of the Gauss curvature and we let $\nabla^{(\Sigma)}$
denote the gradient parallel with $\Sigma$.

\begin{align}
  \mbox{\bf Assumption 2':} & \qquad \qquad \qquad\qquad \min_{\Gamma} 
|\nabla^{(\Sigma)} K|  >0 \; .
\qquad\qquad \qquad\qquad\qquad 
\label{ass2'} \\
\mbox{\bf Assumption 3':} & \qquad \qquad 
\sup_{\om\in S^2} \mbox{card} \{ p\in \Sigma\; : \; \nu(p) =\om\}  <\infty\; .
\qquad\qquad 
\label{ass3'} 
\end{align}
{\bf Assumption 4':}
{ \it 
The set of tangential points on $\Gamma$,
$ \cT= \{ p\in \Gamma \; : \; Z(p)\times w(p)=0\} $,
is finite.  There exists
a positive constant $c>0$ such 
that for any $p\in\Gamma$
\be
    | Z(p) \times w(p)|
  \ge c \cdot d(p)\; ,
\label{PAPlo'}
\ee
where $w$ is the  unit tangent vector of $\Gamma$,
 $Z$ is the unit vector in the principal direction of zero curvature
along $\Gamma$
and for any $p\in \Sigma$ 
$$
 d(p)= \left\{ \begin{array}{ll} 
  1 \quad &\mbox{if }   \cT =\emptyset \\
  \mbox{\rm dist }(p, \cT) \quad &\mbox{if }   \cT \neq \emptyset  \; .
\end{array}
\right.
$$
Moreover, if $\partial\Sigma\neq \emptyset$, we also 
assume that  $\partial\Sigma$ is transversal to $\Gamma_a$:
\be
p\in \Gamma\cap\partial\Sigma \; , \; u\in T_p(\partial\Sigma)
\; \Longrightarrow \; 
  |u \times w(p)| \ge 
 c\| u\|  \; ,
\label{ass4boundary1}
\ee 
and
\be
p\in \Gamma\cap\partial\Sigma \; \Longrightarrow \; 
  |Z(p) \times w(p)| \ge 
 c  \; .
\label{ass4boundary}
\ee
}

\medskip\noindent
Similarly to the proof of Theorem \ref{prop:noc} and Corollary
\ref{cor:L4},  we obtain:

\begin{theorem}\label{cor:sigma}
Let the smooth, compact surface $\Sigma\subset\bR^3$
satisfy Assumptions 2'--4'
and let $f\in C_0^\infty(\Sigma)$
(recall that if $\Sigma$ has a boundary, this means that $f$ 
is supported away from $\partial\Sigma$).
Then the Fourier transform $\wh\mu(\xi)$
of the measure $\rd\mu= f\rd m$ on $\Sigma$ satisfies
the bounds  \eqref{fourdec}, \eqref{fourdecbeta},
where $D_a(\om)$ is replaced
with  $D(\om)= \min \{ |\nu(p)\times \om|\; : \; p\in \cT\}$
if $\cT\neq\emptyset$ and $D(\om)=1$ otherwise.
Furthermore, the norm and integral bounds \eqref{l4}, \eqref{4b}
 hold for $\wh\mu(\xi)$.
The positive constants 
in \eqref{fourdec} and \eqref{l4} depend on $\Sigma$ and $f$.
$\;\;\Box$
\end{theorem}

{\it Remark 1:} 
We note that  Assumptions 2'-- 4' are
generic in the following sense. 
Any two-dimensional surface can be 
locally characterized by its Gauss curvature, i.e.,
by a smooth function $K(p)$. The functions $K$
that violate these assumptions
form a nowhere dense set in the
 $C^\infty$-topology of the curvature function.
For example, violating Assumption 2' would mean that $K$ and
$\nabla K$ simultaneously vanish, which is a non-generic
condition for real-valued functions of two variables.
It would be interesting to replace 
Assumption 2' 
 with the condition  that $\nabla^{(\Sigma)}K$ may vanish
at finitely many points on $\Gamma$ but the vanishing
is of first order. 
In particular, it would allow that the intersection of
the zero sets of the curvatures,
$\{ \kappa_1 = 0\}\cap\{ \kappa_2 =0\}$, is a finite set 
with all intersections are transversal.

{\it Remark 2:} The conditions \eqref{ass4boundary}, \eqref{ass4boundary1}
can always be guaranteed by possibly removing a small tubular
neighborhood of $\partial\Sigma_a$ from $\Sigma_a$. 
Since $f$ is compactly supported away from $\partial\Sigma_a$,
this modification does  not affect the integral
\eqref{whmu}. These conditions ensure that the presence of the boundary
does not affect the proof in Section \ref{sec:proofthm}.

\subsection{The Four Denominator Estimate}\label{sec:4denom}

To formulate the suitable estimate on the integral of \eqref{factor},
we introduce a few notations.
The discrete Laplace operator on $\ell^2(\bZ^3)$
is defined by
$$
   (\Delta f)(x) =  6f(x)-\sum_{|e|=1} f(x+e) \; ,  \qquad
  f\in \ell^2(\bZ^3) \; .
$$
In the Fourier representation, $\Delta$ acts as the multiplication operator
$$
      \wh {( \Delta f)} (p) = -2e(p) \wh f(p)
$$
with
$$
    \wh f(p)= \sum_{x\in \bZ^3} e^{-ip\cdot x} f(x), \qquad 
p=(p_1, p_2, p_3)\in \Tor^3=[-\pi, \pi]^3
$$
and
\be
 e(p)= \sum_{j=1}^3 [ \, 1-\cos p_j \, ] \; .
\label{edef}
\ee
In the physics literature, 
the multiplier $e(p)$ is called the {\it dispersion relation}
of the Laplace operator.

\medskip
\noindent
For any $\a\in\bR$, $u\in \Tor^3$ and $\eta > 0$ we define
\be
    I_{\a, \eta}(u)= 
 \int_{(\Tor^3)^3}
 \frac{\rd p\rd q\rd r}{|\a -e(p)+i\eta||\a -e(q)+i\eta| 
|\a -e(r)+i\eta||\a -e(p+q+r-u)+i\eta|} \; .
\label{defI}
\ee
Our goal is to estimate $I_{\a, \eta}(u)$ for small $\eta$
uniformly in $u$.
Note that the integrand is the absolute value of \eqref{factor}
with $u=0$ since $e(q)=e(-q)$.
 The general case $u\neq 0$ is needed for technical reasons
\cite{ESYdisc}.

For very small $\eta$, 
the integrand in \eqref{defI} is almost singular 
on the level sets $\a = e(p)$, $\a = e(q)$, $\a= e(r)$
and $\a = e(p+q+r-u)$ in the space $(p,q,r)\in (\Tor^3)^3$, 
and the main contribution to the integral comes from the
the intersection of   small neighborhoods of these level sets.
%concentrated 
%on the intersection of   small neighborhoods
%of the level sets $\a = e(p)$, $\a = e(q)$, $\a= e(r)$
%and $\a = e(p+q+r-u)$ in the space $(p,q,r)\in (\Tor^3)^3$. 
We assume that $\a$ is away from
the critical values of $e(p)$, i.e. away from 0, 2, 4, 6.
This  guarantees that the $\a$-level sets are locally
embedded in a regular foliation of neighboring level sets.

For any real number $\a$ we define
\be
    \tri \a \tri = \min \Big\{ |\a|, |\a-2|, |\a -3|, |\a-4|, |\a-6|
 \Big\} \; ,
\label{tripledef}
\ee
and we will assume that $\tri a \tri$ is bounded away from 0.
The value 3 is not a critical value of $e(p)$ but the level
set $e(p)=3$ has a different type of degeneracy that needs to be avoided
(flat umbilic points, see later).

\begin{theorem}[Four Denominator Estimate]\label{thm:4de}
 Let $0<\eta\leq \frac{1}{2}$.
For any $\Lambda>\eta$ there exists a positive constant $C_\Lambda$
such that for any $\a$ with $\tri \a\tri \ge \Lambda$ 
\be
   \sup_{u\in \Tor^3} I_{\a, \eta}(u) \leq C_\Lambda |\log\eta|^{14} \; .
\label{4den}
\ee
\end{theorem}

{\it Remark.}
If we estimated one of the denominators 
in \eqref{defI} by the trivial $\eta^{-1}$
supremum bound, then the remaining three denominators could
independently be integrated out by using the  fairly straightforward
bound (the proof will be given in Section \ref{sec:tech}):
\begin{lemma}\label{lemma:log}
There exists a constant $C$ such that for any $0<\eta \leq \frac{1}{2}$
\be
   \sup_{\a\in \bR}
 \int_{\Tor^3} \frac{\rd p}{|\a - e(p)+ i\eta|} \leq C |\log\eta| \;  .
\label{logeta}
\ee
\end{lemma}
With this lemma, we
 thus would directly obtain $ I_{\a, \eta}(u) \leq C\eta^{-1} |\log\eta|^3$.

The bound \eqref{4den} is  a significant improvement
over this trivial estimate. In particular it shows
that the recollision terms are negligible in the perturbation
expansion (see \cite{ESYdisc}  for more details).
 This is one of the key technical
results behind the proof of the quantum diffusion
of the random Schr\"odinger evolution on the cubic lattice.

\subsection{Remarks on continuous vs. discrete case}

\begin{figure}
\begin{center}
\epsfig{file=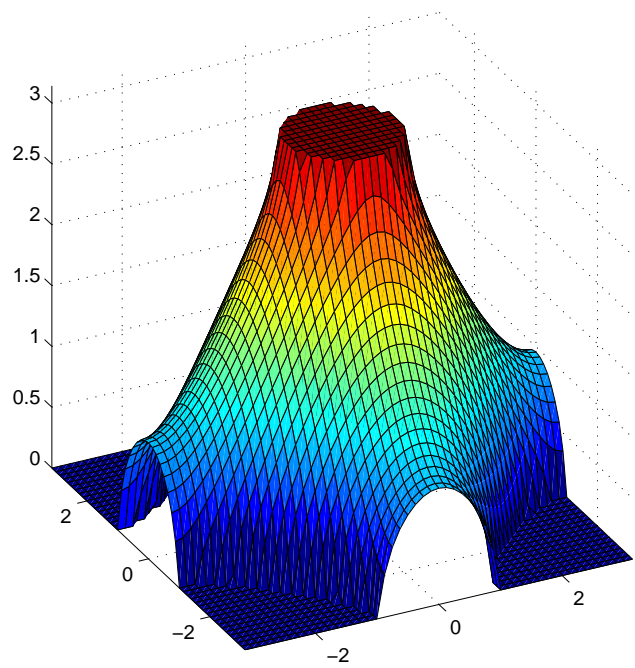,scale=.75}
\end{center}
\caption{Level set of $e(p)=\alpha$ for $2<\a<4$}\label{fig:levelset}
\end{figure}

Let us briefly discuss the relation between the continuous and the lattice case
in the random Schr\" odinger problem. 
The Four Denominator Estimate is used to control
the recollision Feynman diagrams,
as discussed in Section \ref{sec:motiv}, 
for the discrete random Schr\"odinger
evolution \cite{ESYdisc}.
The same  diagrams were estimated in the 
proof of the continuum model as well, \cite{ESYI}--\cite{ESYII}.
 The fundamental
difference is that the level sets of the
continuum dispersion relation, $e_{c}(p) = \frac{1}{2} p^2$,
$p\in \bR^3$, are uniformly convex surfaces (spheres).
The level sets $\Sigma_\a=\{ p\; : \; \a = e(p)\}\subset \Tor^3$
of the discrete dispersion relation \eqref{edef}
are uniformly convex only for $\a\in (0, 2)\cup (4, 6)$.
For $\a\in (2,4)$, the level surfaces $\Sigma_\a$ are not convex,
their Gauss curvature
vanishes along a one-dimensional submanifold and
they even contain straight lines 
(see Figure \ref{fig:levelset}).

In the continuum model, 
the uniform convexity 
implies that a level set  $\{ p \; : \; \a = e_c(p)\} $
and its shifted copy $\{ p \; : \; \a = e_c(p+q) \}$
have transversal intersection or they touch each
other only at a point. This geometric fact
is the key behind the {\it Two Denominator Estimate}
for the continuous dispersion relation $e_c(p)=\frac{1}{2}p^2$
(see Lemma A.1 in \cite{ESYI}):
\be
    \sup_\a
  \int_{\bR^3} \frac{\rd \mu(p)}{|\a - e_c(p)+ i\eta|
 \, |\a - e_c(p+q) + i\eta|}
   \leq \frac{C|\log\eta|^2}{\tri q \tri_\eta}
\label{2den}
\ee
with $\tri q\tri_\eta = \eta + \min \{ |q|, 1\}$
and $\rd\mu$ compactly supported.
In particular, this estimate gives a short proof of 
the continuous version of the four-denominator estimate
with a bound $C|\log\eta|^4$, since
it  allows one to eliminate two denominators
by integrating out one free variable.
% at the expense of collecting
%a point-like singularity $\tri q \tri_\eta^{-1}$.
The other two denominators can then be easily integrated out
using \eqref{logeta}.
%by using the remaining two integration variables
%and collecting
% additional $|\log\eta|$-factors,
%similarly to the bound \eqref{logeta}.  The point
%singularity $\tri q \tri_\eta^{-1}$ causes no essential complication
%(see (A.7) of \cite{ESYI}).

A similar short proof of the four-denominator
bound in the lattice case
is not possible since the analogue of \eqref{2den}
does not hold for $\a\in (2,4)$.
It is easy to see that the two-denominator
 integral ( \eqref{2den} with $e(p)$ instead of $e_c(p)$)
 may be of order $\eta^{-1/2}$
if $q$ shifts along one of the straight line segments contained in 
$\Sigma_\a$. Actually,   only a weaker  upper bound of
order $\eta^{-3/4}$ was proven in \cite{ESYdisc}. (T. Chen
also proved in \cite{Ch} 
 a somewhat weaker three-denominator bound of order $\eta^{-4/5}$.)
 This 
bound is not sufficient to conclude the 
estimate of the recollision term 
for the lattice model in the same way as it was done
for the continuous case in \cite{ESYI}--\cite{ESYII}.

\section{Proof of Theorem \ref{prop:noc}}\label{sec:proofthm}
\setcounter{equation}{0}

In this section we fix $a\in \cI$ and we will work on
the surface $\Sigma_a$. We will define various quantities
that depend on $a$. We will usually omit the dependence
on $a$ in the notation, i.e. we write $\Sigma$, $\Gamma$, $p^{(j)}$, $N$,
$d(p)$, $D(\zeta)$ etc. instead of $\Sigma_a$, $\Gamma_a$, $p^{(j)}_a$, $N_a$,
$d_a(p)$, $D_a(\zeta)$,
but all estimates and constants
will be uniform in $a\in \cI$. 
Let $ \Omega=\{ C_0,  C_1, \ldots,  C_6\}$ be the 
set of constants 
from \eqref{C1def} and 
Assumptions 1--4. Large or small positive
constants depending on $\Omega$ 
 will be denoted by $C_\Omega\gg 1$ or $0<c_\Omega\ll 1$
whose values may change from line to line.
The notation $A\sim_\Omega B$ will refer to comparability
up to $\Omega$-dependent positive
constants, $c_\Omega \leq A/B \leq C_\Omega$.
The notation $\sim$
will be used for comparability up to a universal constant.

%\subsection{Geometry in the neighborhood of $\{ K=0\}$.}
\subsection{Geometry at small Gauss curvature}
%$\{ K=0\}$.}

We  recall that the zero set $\Gamma=\ZK \cap \Sigma$
is a finite union of disjoint regular curves. By \eqref{ass2}
there exists a small constant
$c_0$, depending on $C_1, C_2, C_3$,
 such that for each $a\in\cI$ and all $ |\varrho|\leq c_0$
the sets 
$$
 \Gamma(\varrho)= \Sigma \cap \ZK(\varrho), 
\qquad
\ZK(\varrho) = \{ p \in \cD : K(p) = \rho\}
$$
are regular curves and they form a foliation 
in the tubular neighborhood
\be
   \cN = \{ p\in \Sigma \; : \; |K(p)|< c_0\}
\label{cndef}
\ee
of the curves $\Gamma$ on the surface
$\Sigma$.
Moreover, by \eqref{kappasum} we can choose $c_0$ so small that
on each connected component of $\cN$ one of the curvatures 
is much smaller than the other one. 
Then the principal curvatures and the principal curvature directions 
depend smoothly on $p\in\cN$ with uniform bounds on the derivatives.
We will work on one of these components that we continue to denote by $\cN$,
and for definiteness we assume  $\kappa_1\ll\kappa_2$.

The principal curvature direction of $\kappa_1$ defines a smooth
unit vectorfield in $\cN$ that coincides with  the vectorfield $Z$
on $\Gamma$ and hence it will also be denoted by $Z$. 

Recall that by  Assumption 4, there is a tangency of the integral 
curves of $Z$ and $W$ only at finitely many points, at most $N$ of them. 
We will present the proofs for the case $N\ge1$. 
The case $N=0$ is much easier since these two foliations
are uniformly transversal. We will not discuss
this case in detail, but the statements made below 
remain valid. 

Since $Z$ changes linearly in the neighborhood of the tangential
points (Assumption 4) and $Z$ is regular, 
 the points $\{ p^{(j)}_a\; : j=1, \ldots , N_a\}$
are separated from each other, i.e.
\be
      c_1=\min_{a\in \cI} \min_{j\neq k} |p^{(j)}_a-p^{(k)}_a|  >0
\label{c1}
\ee
and $c_1$ is bounded from below by a positive, $\Omega$-dependent constant.
The uniformity in $a$  follows from the fact that the angle between
$Z(p)$ and $w(p)$
is a regular function as $p$ moves on $\Gamma=\Gamma_a$, in particular
its second derivative is bounded. Since near to a
tangential point $p^{(j)}=p^{(j)}_a$ this angle increases
linearly at a positive speed at least $C_6>0$ (uniformly in $a$),
it cannot turn back to zero before $p$ moved at least
a positive distance away from $p^{(j)}_a$.
This shows the lower bound \eqref{c1}.

\medskip
\noindent

\begin{figure}
\begin{center}
\epsfig{file=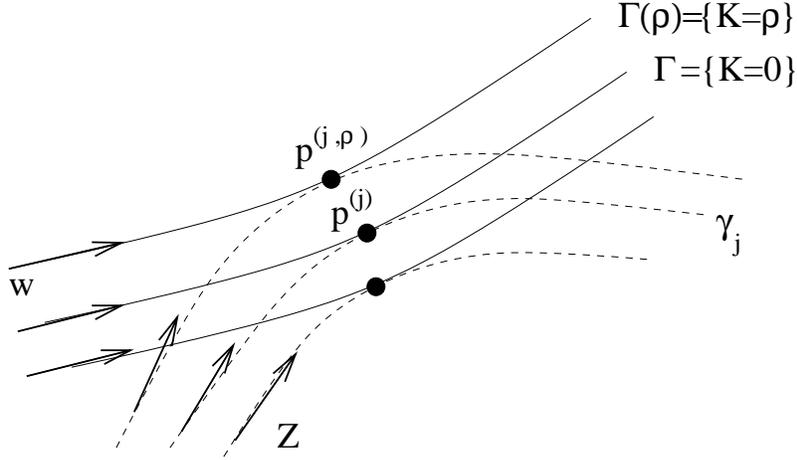,scale=.75}
\end{center}
\caption{Foliations $\Gamma$ and $\gamma$ with their first order tangencies}
\label{fig:gamma}
\end{figure}

The curves $\Gamma(\varrho)$,
$|\varrho|\leq c_0$, form a regular  foliation
of $\cN$ and $\Gamma$ is embedded in this foliation.
The unit tangent vector to this foliation is
 $w(p)$ defined in \eqref{wdef}.
For any point  $p\in \cN$,
let $\gamma_p\subset \cN$ 
be the integral curve of $Z$ that goes through $p$.
If $p^{(j)}$ is one of the points on $\Gamma$
from Assumption 4, then $\gamma_j=\gamma_{p^{(j)}}$
is tangent to $\Gamma$ at $p^{(j)}$, but their curvatures
differ by the linear lower bound \eqref{ass4},
 i.e. the tangency of these two curves 
is precisely of first order (see Fig. \ref{fig:gamma}).

 The following lemma
gives a lower bound on the transversality of the
foliations $\gamma$ and $\Gamma$: 
\begin{lemma}\label{lemma:trans}
For a sufficiently small $c_0$, depending only on $\Omega$, we have
\be
    p\in \cN \;  \; \Longrightarrow \; 
  |Z(p) \times w(p)  |
\ge c_\Omega \cdot d(p) -
 C_\Omega |K(p)| \; .
\label{transversal}
\ee
\end{lemma}

{\it Proof.} If $N=0$, then \eqref{transversal} follows
directly from  \eqref{ass4} by regularity
and $d(p)=1$ assuming $c_0$ is sufficiently small. Thus we
can assume $N\ge 1$.
Due to the regularity of the foliations $\gamma$ and $\Gamma=\{ 
\Gamma(\varrho) \; : \; |\varrho|\leq c_0\}$
and due to their different curvatures at the tangential points,
these two foliations can be
 mapped by a regular bijection $\Phi$
from the neighborhood of each tangential point in $\Sigma$ into
the foliations $\{ v = $ const $ \}$ and $\{ v = u^2 \}$
in the $(u,v)\in \bR^2$ plane near the origin.

Translating this picture into the $\gamma$ and  $\Gamma$
foliations on $\Sigma$,
this means that, for small enough $c_0$, 
there exist tangential points
$p^{(j, \varrho)}\in \Gamma(\varrho)$,
where the curves $\gamma_{j, \varrho}= \gamma_{p^{(j,\varrho)}}$
and $\Gamma(\varrho)$ have first order tangencies
for any $|\varrho|\leq c_0$. For any $p\in\cN$,
we let 
$$
d^{(\varrho)}(p)= \min \{ |p-p^{(j,\varrho)}|\;  : \; 
1\leq j \leq N\},
$$
 where $\varrho$ is uniquely defined by $\varrho=K(p)$.
Moreover, by the regularity of $\Phi$,
 the  tangential points $p^{(j,\varrho)}$
 are $C^1$ functions of $\varrho$ with bounded derivatives. 
In particular,
\be
 \big| 
  d^{(\varrho)}(p) - d(p)\big|\leq C_\Omega |\varrho|,
\quad \varrho=K(p),
\label{lip}
\ee
uniformly in $p\in\cN$, $a\in \cI$, if $c_0$ is sufficiently small.

The regular bijection $\Phi$ also guarantees that
the foliations $\gamma$ and $\Gamma$ have first order tangencies
near the tangential points $p^{(j,\varrho)}$
on each $\Gamma(\varrho)$.
For a sufficiently small
$\Omega$-dependent positive constant $c_2\leq \frac{1}{2}c_1$
and by possibly reducing $c_0$  we thus have 
\be
  |p-p^{(j,\varrho)}|\leq c_2 \;\; 
 \;\; \Longrightarrow \;\; 
\quad  |Z(p) \times w(p)  |\ge
c_\Omega |p-p^{(j,\varrho)}|
\label{2cur2}
\ee
for any $p\in \cN$, $a\in\cI$, $1\leq j\leq N$, where $\varrho=K(p)$.

Away from the tangential points, we first use
$$
     p\in \Gamma \;, 
\;  d(p)\ge \frac{1}{2} \; c_2  \;  \; \Longrightarrow \; 
  |Z(p) \times w(p)  |\ge c_\Omega
$$
by compactness and the continuity 
of the function $|Z(p) \times w(p)  |$ on $\Gamma$
with zeros $p^{(j)}$.
Then we extend this lower bound for $|\varrho|\leq c_0$
by continuity if $c_0$ is sufficiently small:
\be 
  p\in \Gamma(\varrho) \;, 
\;  d_{a,\varrho}(p)\ge  c_2  \;  \; \Longrightarrow \; 
  |Z(p) \times w(p)  |\ge c_\Omega \; .
\label{away}
\ee
By combining this uniform bound with the estimate \eqref{2cur2} near
the tangential
points and by using \eqref{lip}, we have
\be
    p\in \Gamma(\varrho), \;  \; \Longrightarrow \; 
  |Z(p) \times w(p)  |\ge
c_\Omega \cdot d^{(\varrho)}(p)  \quad\mbox{with} \;\;\varrho= K(p)
\label{trans2}
\ee
By using \eqref{lip}, we obtain \eqref{transversal}. $\;\;\Box$

\medskip
\noindent

\begin{figure}
\begin{center}
\epsfig{file=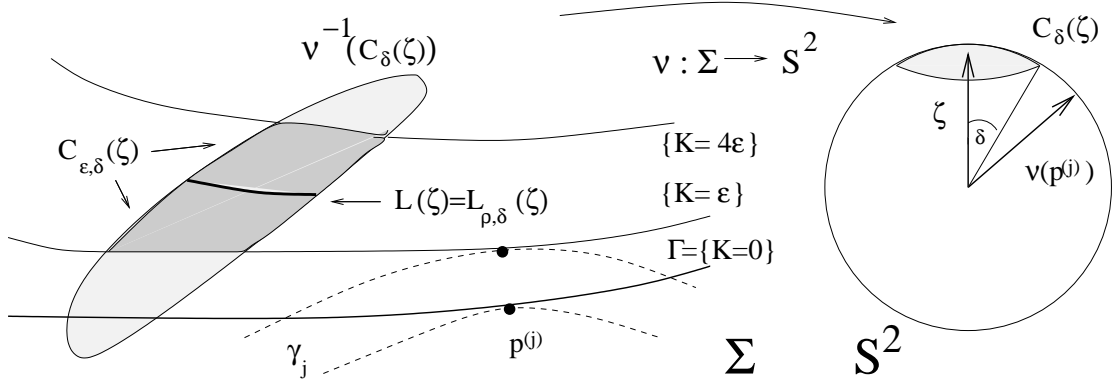,scale=.60}
\end{center}
\caption{Intersection of $\e \leq K \leq 4\e$ and the preimage
of the spherical cap $C_\delta(\zeta)$}
\label{fig:L}
\end{figure}

In the stationary phase analysis we will have to estimate 
the  volume of a regime $\{ q\in \Sigma \; : \; K(q)\sim \e\}$ intersected
with the preimage  of a small spherical
cap $C_\delta(\zeta)= \{ \om \in S^2\; : \; |\om\times\zeta|\leq \delta\}$ 
around $\zeta\in S^2$
under the Gauss
map $\nu:\Sigma\to S^2$. We thus define the set
$$
   C_{\e, \delta}(\zeta)= \Big\{ q\in \cN \; : \; \e \leq |K(q)|\leq 
  4\e \;, \; 
    |\nu(q)\times \zeta|\leq \delta\Big\} \; .
$$
for any $\zeta\in S^2$ (see Fig. \ref{fig:L}).

\begin{lemma}\label{keylemma}
 Let  $c_0$ be sufficiently small, depending on $\Omega$, let
 $0<\e \leq c_0$ and $\delta>0$. Then for any $\zeta\in S^2$
at least one of the following holds:
\be
   \mbox{either} \quad
      \mbox{\rm vol}_\Sigma \;\big[ C_{\e,\delta}(\zeta) \big]\leq  
\frac{C_\Omega \e\delta}{ \big[ D(\zeta) \big]^{1/2}} \; ,
   \quad \mbox{or} \;\;   D(\zeta) \leq C_\Omega (\e+\delta) \; .
 \label{voles}
\ee

\end{lemma}

{\it Proof.} We will work in one component of $\cN$ and we recall
that $\kappa_1\ll \kappa_2$ in this component, i.e.
$\kappa_2 \ge c_\Omega$ by \eqref{kappasum}
and  $|\kappa_1|\sim_\Omega |K|$.

The principal curvature
direction corresponding to $\kappa_2$ is orthogonal
to the foliation $\gamma$, thus the normal vector $\nu(b)$
changes linearly with a coefficient proportional
to $\kappa_2$ if the base point $b$ is moving transversally
to $\gamma$. 
If $b$ moves along a curve $\gamma$, then the change is
  proportional to $\kappa_1$
(plus a quadratic correction):
\be
     |\nu(b)- \nu(b')|\leq C_\Omega \Big[\;  |b-b'|^2 + \big( \sup_{[b,b']}
 |\kappa_1|] \big) \; |b-b'|\; \Big]
\label{variation}
\ee
if $b'\in \gamma_b$ and the supremum is taken on the curve segment
between $b$ and $b'$.

Let $q\in C_{\e, \delta}(\zeta )$ and let $d(q) = |q-p^{(j)}|$
for an appropriate $j$.
Then the base point $q$
can first be moved transversally with a distance less than
 $C_\Omega |K(q)|$ to reach the curve
$\gamma_j$, then it can be moved along this curve
with a distance less than $C_\Omega d(q)$ to reach $p^{(j)}$.
The motion stays in a  neighborhood of $\Gamma$
of width comparable with $d(q)$ or smaller,  so
the Gauss curvature, and thus $\kappa_1$,
is bounded by $C_\Omega d(q)$ along the whole
motion.

{F}rom  \eqref{variation} we have
\be
|\nu(q)\times \nu(p_j)| \leq 
|\nu(q) - \nu(p_j)| \le
C_\Omega  ( |K(q)| + d(q)^2)
\label{nuq}
\ee
by using $|\nu \times \nu'|\leq  |\nu-\nu'|$
for unit vectors.
Furthermore, $q\in C_{\e,\delta}(\zeta )$ implies that
$|\zeta \times \nu(q)|\leq  \delta$, thus we obtain
\be
  D(\zeta )\leq C_\Omega ( |K(q)| + d(q)^2  +\delta)
\label{Dd}
\ee
since
$|\nu\times \nu''|\leq (|\nu\times \nu'|+|\nu' - \nu''|)$ for
unit vectors.

If there exists a point $q\in C_{\e, \delta}(\zeta)$  with
$d(q)\leq C_\Omega |K(q)|^{1/2}$, then \eqref{Dd}
implies the second statement of \eqref{voles}.
For the rest of the proof we thus can assume that
\be
\forall \; q\in C_{\e,\delta}(\zeta) \;\;\Longrightarrow\;\;
C_\Omega |K(q)|^{1/2}\leq d(q)\;.
\label{kd}
\ee
In particular, by \eqref{Dd},
\be
  D(\zeta )\leq C_\Omega ( d(q)^2  +\delta) \; ,\qquad 
\forall \; q\in C_{\e,\delta}(\zeta)\; .
\label{Dd1}
\ee

\medskip
\noindent

In this case we will show that  
$\mbox{vol} \big[ C_{\e,\delta}(\zeta)\big] 
\leq C_\Omega \e\delta / [D(\zeta)]^{1/2}$.
 For $\e\leq |\varrho|\leq 4\e$, we
define (see Fig. \ref{fig:L})
$$
   L(\zeta)=L_{\varrho, \delta}(\zeta)= \Gamma(\varrho)\cap 
  C_{\e,\delta}(\zeta)=
   \Big\{ q\in \Gamma(\varrho) \; :
 \;  |\nu(q)\times \zeta|\leq \delta\Big\} \; .
$$

\begin{proposition}\label{prop:cL}
Suppose that \eqref{kd} holds.
Then the one-dimensional
measure of $ L(\zeta)$, as a subset of
the curve $\Gamma(\varrho)$, satisfies
\be
   |L(\zeta)|\leq  
\frac{C_\Omega \delta}{ \big[D(\zeta)\big]^{1/2}}
\label{lengthes}
\ee
for any $\e\leq |\varrho|\leq 4\e$.
\end{proposition}

By this Proposition,
 the first statement in \eqref{voles}
 follows by integration over $\varrho\in [\e, 4\e]$
and by the regularity of the foliation $\Gamma$. 
This completes the proof of Lemma \ref{keylemma}. $\;\;\Box$.

\begin{figure}
\begin{center}
\epsfig{file=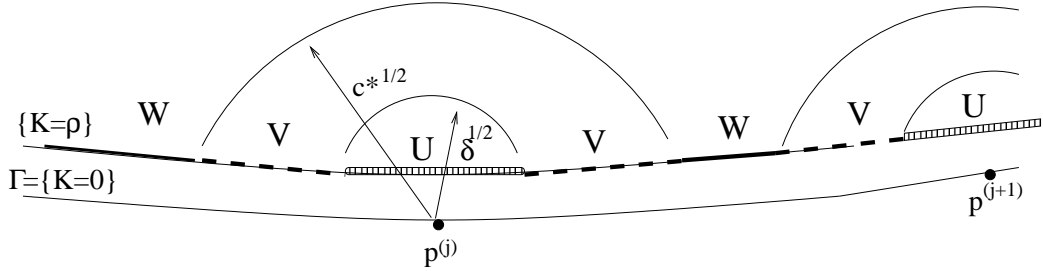,scale=.60}
\end{center}
\caption{The subsets $U, V, W$ on the line $\Gamma_\varrho= \{ K = \varrho\}$}
\label{fig:uvw}
\end{figure}

\medskip
\noindent
{\it Proof of Proposition \ref{prop:cL}.}
We can  assume
that $\delta$ is small, otherwise \eqref{lengthes} follows
from the boundedness of $|\Gamma(\varrho)|$ and $D(\zeta)$.
We now fix $\varrho\in\pm [\e,4\e]$ and
define the set
$$ 
    W=    W_{\varrho}= \Big\{ q\in \Gamma(\varrho) \; : \;
       d(q)^2\ge c^*\Big\} \;,
$$
where $c^*< 1$ is a sufficiently small 
$\Omega$-dependent constant.  By \eqref{c1} and for a sufficiently
small $c_0$ and $c^*$ it is clear that its complement
$W^c=\Gamma(\varrho)\setminus W$
 consists of at most $N$ connected pieces of $\Gamma(\varrho)$
and thus $W$
 consists of at most $N+1$ pieces (in particular, if
$N=0$, then $\Gamma(\varrho)= W$).
Furthermore, we also define
$$
   U=    U_{\varrho,\delta}= \Big\{ q\in \Gamma(\varrho) \; : \;
       d(q)^2\leq \min \{ \delta, c^*\}\Big\} \subset W^c\;,
$$
that also consists of at most $N$ connected pieces. The complement
$$
   V=W^c\setminus U=\Big\{ q\in \Gamma(\varrho) \; : \;
       \min \{ \delta, c^*\}\leq d(q)^2\leq c^* \Big\} 
$$
 thus consists of at most $2N$ connected pieces. The interesting case
is when $\delta < c^*$, i.e. $V\neq\emptyset$ (see Fig. \ref{fig:uvw}).

We decompose 
$$
     L(\zeta)= \big(L(\zeta)\cap U\big)\cup \big( L(\zeta)\cap V\big)
  \cup \big( L(\zeta)\cap W\big)
$$
and estimate the length of each piece separately.

For the first piece, we use the trivial bound
$$
     |L(\zeta)\cap U| \leq |U| \leq C_\Omega N\delta^{1/2}\leq C_\Omega
 \delta^{1/2}\;,
$$
as $U$ consists of at most $N$ pieces of length at most $\sim \delta^{1/2}$
and $N\leq C_\Omega$.
The resulting $C_\Omega\delta^{1/2}$ can be  bounded by
$C_\Omega \delta /D(\zeta)^{1/2}$
since $D(\zeta) \leq C_\Omega \delta$ from \eqref{Dd1}
if $L(\zeta)\cap U\neq\emptyset$.

For the other two pieces, we recall the bound
 $|Z(q) \times w(q)|\ge c_\Omega\cdot d(q) - C_\Omega |K(q)|$
 from \eqref{transversal}.
By the condition (\ref{kd}),  $C_\Omega |K(q)|^{1/2}\leq d(q)$
holds on $L(\zeta)\subset C_{\e,\delta}(\zeta)$, so 
$|Z(q) \times w(q)|\ge c_\Omega\cdot d(q)$
if $c_0$ (hence $|K(q)|$) is sufficiently small. Thus
the transversality angle between the two foliations
is at least $c_\Omega d(q)$, i.e. $\nu(q)$ changes
at least at a rate $\sim_\Omega d(q)$ as $q$ is moving 
along $L(\zeta)$
\be
   q\in L(\zeta)\;\;\; \Longrightarrow
\;\; |\nabla_w \nu(q)|\ge c_\Omega d(q).
\label{tra}
\ee

We need the following elementary lemma:

\begin{lemma}\label{lemma:elem}
i)  Let $I$ be a compact interval and $g: I\to \bR^3$
be a twice differentiable function with $\inf_I |g'| \ge \lambda>0$.
Then for any $\delta$ we have
\be
      \Big|\Big\{ q\in I \; : \; |g(q)|\leq \delta\Big\}\Big| \leq
     8\delta \lambda^{-1}+16|I|\delta \lambda^{-2} \max_I | g''|\; ,
\label{om2}
\ee
where $\big| \cdot \big|$ denotes the one-dimensional Lebesgue measure.

ii) Let $h: I\to \bR$ with $\inf_I |h'| \ge \lambda>0$ and assume
that $h'$ has a definite sign in $I$. Then
\be
      \Big|\Big\{ q\in I \; : \; |h(q)|\leq \delta\Big\}\Big| \leq
     4\delta \lambda^{-1}\; .
\label{om1}
\ee
\end{lemma}

We will
apply the first part of this lemma to the function $g(q)=\nu(q)\times \zeta$
along $\Gamma(\varrho)$
on each connected piece of
 $L(\zeta)\cap W$. Clearly $|g''|\leq C_\Omega$.
Since $\nu$ is a normal vector, 
 its variation  along  $\Gamma(\varrho)$,
$\nabla_w\nu$, is orthogonal to $\nu$. Since $\nu$ and $\zeta$
are almost parallel on $L_{a,\varrho}(\zeta)$ (assuming $\delta\ll 1$),
the variation
of $g$ is comparable with the variation  of $\nu$, i.e.
$|\nabla_w g|\ge \frac{1}{2} |\nabla_w\nu|$.
 On  $L(\zeta)\cap W$ we have  $d(q)^2 \ge c^*$, thus
 $|\nabla_w g|\ge c_\Omega$ from \eqref{tra}.
By \eqref{om2}, we have
$$
  |L(\zeta)\cap W|\leq C_\Omega N\delta \le C_\Omega\delta \; ,
$$
which is smaller than the bound \eqref{lengthes} since
$D(\zeta)\leq C_\Omega$.

\medskip
\noindent
Finally, we consider each connected piece of $L(\zeta)\cap V$.
Let $S$ be one of them.
Let $R(q)\in T_p\Sigma$ be the unit vectorfield orthogonal to $Z(q)$,
i.e. it is the direction of principal curvature belonging to $\kappa_2$
(recall that $|\kappa_2|\ge c_\Omega$ on $\cN$).
We decompose the variation of $\nu$ 
along $\Gamma(\varrho)$ as
\be
    \nabla_w\nu = (\nabla_w\nu\cdot R)R + (\nabla_w\nu\cdot Z)Z 
   = \kappa_2 (R\cdot w)R + \kappa_1(Z\cdot w)Z\;,
\label{deco}
\ee
by using $\nabla_R\nu \cdot R = \kappa_2$, $\nabla_Z\nu \cdot Z = \kappa_1$
and $\nabla_R\nu \cdot Z = \nabla_Z\nu \cdot R =0$.

Within $L(\zeta)$, the $R$-component of $\nabla_w\nu$ is bounded
from below  $|\nabla_w \nu(q)\cdot R(q)|\ = |\kappa_2| |R(q)\cdot w(q)|
\ge c_\Omega d(q)$ by
\eqref{trans2}.
Moreover $\nabla_w \nu\cdot R =\kappa_2 (R\cdot w)$ has a definite sign on $S$.
The $Z$ component of  $\nabla_w\nu$ is bounded by
\be
   |\nabla_w\nu(q)\cdot Z(q)|\leq C_\Omega |K(q)|\leq C_\Omega d(q)^2
\label{zw}
\ee
by using $|\kappa_1|\leq C_\Omega |K|$ and \eqref{kd}. 

Fix a point $q_0\in S$ and define $h(q)= R(q_0)\cdot (\nu(q)-\zeta)$.
Its derivative along $w$ is given by
\be
   h'(q) =  (\nabla_w\nu(q)\cdot R(q)) (R(q)\cdot R(q_0))
   + (\nabla_w\nu(q)\cdot Z(q)) (Z(q)\cdot R(q_0))
\label{hprime}
\ee
by using \eqref{deco}. For a  sufficiently small $c^*$, the vectorfield $R$
does not change much on $S$, thus $R(q)\cdot R(q_0)\ge \frac{1}{2}$
for all $q\in S$. Thus the first term in \eqref{hprime} has a definite
sign and it is bigger than $c_\Omega d(q)$ in absolute value.
The second term is smaller than $C_\Omega d(q)^2$. 
For a sufficiently small $c^*$ we thus have $|h'(q)|\ge c_\Omega d(q)$
and $h'$ has a definite sign.
Moreover, on $V$,  $d(q)^2 \ge 2\delta$ implies 
 $d(q)^2\ge c_\Omega D(\zeta)$, see \eqref{Dd1}.
Thus for the function $h(q)$,
defined on the connected piece $S$, it holds that
$|h'|\ge c_\Omega [ D(\zeta)]^{1/2}$
and $h'$ has definite sign.

Since $|\nu(q) \times \zeta| 
\ge c |\nu(q)-\zeta| \ge c R(q_0)\cdot (\nu(q)-\zeta)=ch(q)$,
we can directly apply \eqref{om1} for each connected piece $S$
to obtain
$$
    |L(\zeta)\cap V|\leq 
\frac{C_\Omega N \delta}{\big[ D(\zeta)\big]^{1/2}}\;,
$$
and the proof of Proposition \ref{prop:cL} is complete. $\;\;\Box$

\medskip
\noindent
{\it Proof  of Lemma \ref{lemma:elem}.}  We denote $M=\max_I | g''|$.
Let $q_0\in I$ be a point with $|g(q_0)|\leq \delta$. By Taylor expansion,
$$
      |g(q)| \ge |g'(q_0)||q-q_0| - |g(q_0)| - M    |q-q_0|^2 \; .
$$
i.e. $|g(q)| >\delta$ if $ \frac{4\delta}{\lambda}
  <|q-q_0|\leq \frac{\lambda}{2M}$.
Thus within the interval
 $q\in[q_0- \frac{\lambda}{2M}, q_0+\frac{\lambda}{2M}]
\cap I$
only a subset of measure at most $\frac{8\delta}{\lambda}$ can
satisfy $|g(q)|\leq \delta$, i.e. the density of
the set $\{ |g|\leq \delta\}$ is not bigger than
$\max \{16\delta M\lambda^{-2}, 8\delta/(\lambda |I|)\}$. 
This proves \eqref{om2}.
The proof of \eqref{om1} is similar, by noticing that
$|h(q)|>\delta$ if $2\delta/\lambda\leq |q-q_0|$.
 $\;\;\Box$

\subsection{Dyadic decomposition}

For any vector $\xi\in \bR^3$, let
 $\xi =r\om_\xi$  be its polar decomposition
 with $r=|\xi|$, $\om_\xi\in S^2$.  We will estimate $\wh\mu_a(\xi)$
and we will omit $a$ from the notation as before.

\medskip
\noindent
We recall the definition of $\cN$ from \eqref{cndef} and
we assume that $c_0$ is so small as required
in Lemma \ref{lemma:trans} and \ref{keylemma}.
Let $S_0= \Sigma\setminus \cN$
 be the complement of this neighborhood.
Let $1\leq k \leq L$ be an integer and set
\be
     S_k= \{ p\in \Sigma\; :\;  2^{-k}c_0
\leq |K(p)|\leq 2^{-k+2}c_0\} \; .
\label{Sk}
\ee
We also set
$$
    S_\infty= \{ p\in \Sigma\; :\;   |K(p)|\leq 2^{-L+1}c_0\} \; ,
$$
then $S_0, S_1, \ldots , S_\infty$ cover $\Sigma$ with
 overlaps.
For the two-dimensional (surface) volume of these sets, we
clearly have 
$$
\mbox{vol}(S_k)
\sim_\Omega 2^{-k}, \quad k\leq L,\quad \mbox{and}\quad
\mbox{vol}(S_\infty)\leq C_\Omega\cdot 2^{-L} \; .
 $$

We will say that two domains are {\it regular bijective images} of each other
if there is a diffeomorphism $\Phi$ between them
such that the derivatives of $\Phi$ and $\Phi^{-1}$ are both
bounded with a $\Omega$-dependent uniform constant.

Since each $S_k$, $k\ge 1$, is the difference of level sets of 
the regularly foliating  function $K(p)$,
it is a regular bijective
image  of finitely many elongated
rectangles with side-lengths $2^{-k}\times 1$. 
Similarly, $S_0$ can be written as a complement of regular images
of finitely many rectangles. Therefore there exists a
partition of unity, $\psi_0, \psi_1, \ldots, \psi_{L}, \psi_\infty $,
such that $\sum_{k=0}^{L}\psi_k + \psi_\infty \equiv 1$,
$0\leq \psi_k \leq 1$, and
\be
\mbox{supp}\;\psi_\infty \subset S_\infty,\quad
\mbox{supp}\; \psi_k \subset S_k \qquad
|\partial^{\alpha}
\psi_k|\leq C_{\Omega,\alpha} 2^{|\alpha|k}, \quad 0\leq k\leq L
\label{psid}
\ee 
for any multiindex $\alpha$ and $\psi_\infty$ satisfies the same
bounds as $\psi_L$. The constants $C_{\Omega, \a}$ depend on
the $\a$ and $\Omega$.

We split the integral defining $\wh\mu$ as follows
\be
   \wh\mu(\xi) = I_\infty(\xi) + I_0(\xi)+\sum_{k=1}^{L} I_k(\xi)
\label{split}
\ee
with
$$
I_k(\xi)=\int_{\Sigma} e^{i\xi\cdot p}\;
   \psi_k(p) f(p) \rd m(p) \; .
$$
By the estimate on $\mbox{vol}(S_\infty)$, we obtain
\be
  \sup_{\xi\in \bR^3} |I_\infty(\xi)|\leq C^* \cdot 2^{-L} \; 
\label{Iinfty}
\ee
(recall that $C^*$ denotes a constant depending on $\Omega$ and $f$).

To estimate
$\sum_{k=1}^L I_k(\xi)$,
we define
\be
    R_j =\{ \om\in S^2\; : \; 2^{-j}\leq |\om\times\om_\xi|\leq 2^{-j+2}\}
\label{def:R}
\ee
for $j=0,1,2,\ldots $. Notice that $R_j$ consists of two antipodal
spherical annuli with inner radius and width comparable with $2^{-j}$
and $R_j$ lies in two antipodal spherical caps:
$R_j\subset C_j^+\cup C_j^-$ with
$$
    C_j^\pm= \{ \om\in S^2\; : \; 2^{-j-1}\leq
  |\om \mp \om_\xi|\leq 2^{-j+3}\} \; .
$$

We define a partition of unity $0\leq \varphi_j\leq 1$ on  $S^2\setminus
\{ \pm \om_\xi\}$
such that 
\be
 \sum_{j=0}^\infty \varphi_j \equiv 1, 
\qquad \mbox{supp}\; \varphi_j \subset R_j, \qquad |\partial^\alpha\varphi_j|
  \leq C_\alpha \cdot 2^{|\alpha|j}  
\label{phid}
\ee
for any multiindex $\a$, 
and we write, for any $1\leq k\leq L$,  
\be
     I_k(\xi)=\sum_{j=0}^\infty I_{k,j}(\xi), \qquad
    I_{k,j}(\xi)=\int_{\Sigma} e^{ip\cdot\xi} \;
   \psi_k(p)\varphi_j(\nu(p))f(p)\rd m(p) \; .
\label{ikj}
\ee
Let
$$  
   U_{k,j}= S_k\cap \nu^{-1}(R_j)\; ,
$$
where $\nu:\Sigma\to S^2$ is the normal map \eqref{def:nor}.
Then the integration domain for $I_{k,j}$ is contained in $U_{k,j}$,
$$
 \mbox{supp} (\psi_k)\cap \mbox{supp}(\varphi_j\circ \nu)\subset U_{k,j} \; .
$$

By a trivial supremum bound we have
\be
    | I_{k,j}(\xi)| \leq C^*\cdot
     \mbox{vol} ( U_{k,j}) \; ,
 \label{Isup}
\ee
in particular
\be
    | I_{k,j}(\xi)| \leq C^*\; .
\label{trivvi}
\ee

\subsection{A stationary phase lemma}

We need the following Lemma:

\begin{lemma}\label{lemma:stt}
Let $Q$ be a smooth function supported in a sufficiently
small subset $U\subset \Sigma$ 
 so that $|\partial e(p)/\partial p_3| \ge c_\Omega$
holds for $p\in U$.
Suppose that on this neighborhood $|\nu(p) \times\om_\xi|\ge \delta>0$.
Then
\be
    \Bigg| \int_{\Sigma} e^{ip\cdot \xi} Q(p)\rd m(p) \Bigg|
    \leq \frac{C_\Omega}{|\xi|} \cdot \mbox{\rm vol}_\Sigma (U)
  \cdot \Big( \delta^{-2} \| Q \|_\infty+ \delta^{-1}\| Q\|_{C^1} \Big)\; .
\label{statt}
\ee
\end{lemma}

{\it Proof of Lemma \ref{lemma:stt}.}  On the subset $U$,
the surface $\Sigma$ can be coordinatized 
by $p_1, p_2$, i.e.  $U$ embedded in $\bR^3$ or $\Tor^3$ can be described
as a regular function  $p_3=p_3(p_1, p_2)$ 
  with uniformly bounded derivatives.
After a change of variables,
we have (with $\om_\xi= (\om_{\xi,1}, \om_{\xi,2}, \om_{\xi, 3})$)
\be
  \int_{\Sigma} e^{ip\cdot \xi} Q(p)\rd m(p) =\int_{\pi(U)}
   e^{i|\xi |(p_1\om_{\xi,1}+ p_2\om_{\xi,2}+ p_3(p_1, p_2)\om_{\xi,3})}
   \wt Q(p_1, p_2) \rd p_1 \rd p_2
\label{Dphase}
\ee
where $\wt Q(p_1, p_2) = Q(p)(1+|\nabla p_3|^2)^{1/2}$
is evaluated at $p=(p_1, p_2, p_3(p_1, p_2))$
and $\pi(U)$ is the projection of $U$ 
onto the $(p_1, p_2)$-plane. The Jacobian can be computed as
$$
 (1+|\nabla p_3|^2)^{1/2} = |\nabla e|\, 
\Big|\frac{\partial e}{\partial p_3}\Big|^{-1} \; ,
$$
by differentiating the defining equation
 $e(p_1, p_2, p_3(p_1,p_2))=a$,
\be
    \frac{\partial p_3}{\partial p_1} =- \frac{\nu_1(p)}{\nu_3(p)},
    \qquad 
 \frac{\partial p_3}{\partial p_2} = -\frac{\nu_2(p)}{\nu_3(p)} \; ,
\label{nude}
\ee
and by using  \eqref{def:nor}.

The gradient of the phase factor in \eqref{Dphase}, as a function of
$(p_1, p_2)$, can be estimated from below
by
$$
  \Big|\nabla_{(p_1,p_2)}(p_1\om_{\xi,1}+ p_2\om_{\xi,2}+
 p_3(p_1, p_2)\om_{\xi,3})\Big|
  \ge \frac{1}{\sqrt 2}
  \Big| \om_{\xi,1} + \frac{\partial p_3}{\partial p_1} \om_{\xi,3}\Big|
 +\frac{1}{\sqrt 2} 
 \Big| \om_{\xi,2} + \frac{\partial p_3}{\partial p_2} \om_{\xi,3}\Big| \; .
$$
For any two  vectors, $\nu,\om\in \bR^3$, with $\nu_3\neq 0$ we have
$$
   |\om\times \nu| \leq \Big| \om_1 - \frac{\nu_1}{\nu_3} \om_3\Big| \,
   \big( |\nu_2| + |\nu_3|\big)
  + \Big| \om_2 - \frac{\nu_2}{\nu_3} \om_3\Big| \,
   \big( |\nu_1| + |\nu_3|\big)\; ,
$$
i.e. 
$$
   |\om_\xi\times \nu(p)| \leq 2\Bigg(
   \Big| \om_{\xi,1} - \frac{\nu_1(p)}{\nu_3(p)} \om_{\xi,3}\Big|
 + \Big| \om_{\xi,2}  -\frac{\nu_2(p)}{\nu_3(p)} \om_{\xi,3}\Big|\Bigg) \; ,
$$
since $\|\nu\|=1$. Therefore, by using \eqref{nude},
the gradient of the phase factor  in \eqref{Dphase}
is bounded 
from below by $ \frac{1}{\sqrt{8}}\, |\om_\xi\times \nu(p)|\, |\xi|\ge 
\frac{1}{\sqrt{8}} \, \delta
|\xi|$.
The estimate (\ref{statt}) then follows by standard integration by parts
by using \eqref{C1def},
\eqref{ass1} and the lower bound on $|\partial e(p)/\partial p_3|$.
$\;\;\Box$

\subsection{The estimate of $I_{k,j}(\xi)$}\label{sec:ikjest}

Applying Lemma \ref{lemma:stt}
to our integral $I_{k,j}(\xi)$, we obtain
\be
     | I_{k,j}(\xi)| \leq \frac{C_\Omega}{|\xi|}\cdot
     \mbox{vol} ( U_{k,j}) \Bigg( 2^{2j} \cdot  
\| f \|_{\infty}+ 
  2^{j} \cdot  
\big\|\psi_k(p)\varphi_j(\nu(p)) f(p)
\big\|_{C^1} \Bigg)
\label{Istat}
\ee
by using that on the integration domain 
$|\nu(p)\times\om_\xi|\ge 
2^{-j}$ by \eqref{def:R}.

{F}rom \eqref{C1def}, \eqref{ass1} 
 and from the 
bounds (\ref{psid}), (\ref{phid}) we have
$$
    \big\| \psi_k(p)\varphi_j(\nu(p))  f(p)
\big\|_{C^1}
\leq C^* \max\{  2^{k}, 2^{j} \}
$$
i.e.
\be
     | I_{k,j}(\xi)| \leq \frac{C^*}{|\xi|}\cdot
     \mbox{vol} ( U_{k,j}) \cdot \max\{ 2^{k+j}, 2^{2j}\} \; .
\label{IIstat}
\ee
Interpolating
it with (\ref{Isup}),  we have
\be
     | I_{k,j}(\xi)| \leq \frac{C^*}{|\xi|^{3/4}}\cdot
\Big[ \max\{  2^{k+j}, 2^{2j} \}\Big]^{3/4}
\cdot     \mbox{vol} ( U_{k,j}) \; .
\label{Icomb}
\ee

The domain $U_{k,j}$ is contained in $\nu^{-1}(R_j)$,
so its volume is bounded by
\be
    \mbox{vol}( U_{k,j})  \leq C_\Omega \cdot 2^k 
 \cdot \mbox{vol} ( R_j) \leq C_\Omega \cdot 2^{k-2j},
\label{kj}
\ee
using that the Jacobian of $\nu^{-1}$
 is   $|K|^{-1}\sim 2^k$ on the support of
$\psi_k$ and the number of preimages is bounded by \eqref{ass3}.
We also have
\be
  \mbox{vol}( U_{k,j}) \leq \mbox{vol}(S_k) \leq C_\Omega
  \cdot 2^{-k}\; .
\label{2k}
\ee
By  (\ref{Isup}) and (\ref{kj}),  we have
\be
   \sum_{k=1}^{L}
   \sum_{j=L}^\infty |I_{k,j}(\xi)|\leq C^*\cdot  2^{-L}\; ,
\label{tail}
\ee
so these terms can be combined
with the bound \eqref{Iinfty} and
 from now on we can assume that $j\leq L$.

\bigskip
\noindent
To estimate $I_{k,j}(\xi)$ for $1\leq k\leq L$, $0\leq j\leq L$, we
use  Lemma \ref{keylemma} 
to estimate $\mbox{vol} (U_{k,j})$.
Clearly $U_{k,j} \subset  C_{\e, \delta}(\om_\xi)$ with the choice 
$\e=2^{-k}c_0$, $\delta= 2^{-j+2}$  and we
 obtain that 
either $\mbox{vol} (U_{k,j}) \leq C_\Omega \cdot  2^{-k-j} 
D(\om_\xi)^{-1/2}$ or 
$D(\om_\xi)\leq C_\Omega (2^{-k}+2^{-j})$.
In the first case, combining this estimate with  \eqref{kj}
and \eqref{2k}, 
we obtain 
\be
     \mbox{vol} (U_{k,j}) \leq C_\Omega  \cdot \min\Bigg\{  2^{-k}\; , \;
\frac{2^{-k-j}}{ \big[ D(\om_\xi) \big]^{1/2}} \; , \; 
\frac{2^{-\frac{3}{2}j}   }{ \big[ D(\om_\xi) \big]^{1/4}}
  \Bigg\} \; ,
 \label{volbound}
\ee
so together with  \eqref{Icomb} and the boundedness of 
$D(\om_\xi)$, we have
\be
|I_{k,j}(\xi)|\leq \frac{C^*}{|\xi|^{3/4} \big[ D(\om_\xi)\big]^{1/2}} \; .
\label{34}
\ee

In the second case we use the trivial estimate 
\eqref{Isup} 
$$
     | I_{k,j}(\xi)|\leq 
    \mbox{vol}(U_{k,j})
 {\bf 1}\big[  D(\om_\xi) \leq
  C_\Omega(2^{-k}+2^{-j}) \; \big] \; ,
$$
where ${\bf 1}[\ldots ]$ is the characteristic function.
We combine it with \eqref{IIstat}
and with the bound $ \mbox{vol}(U_{k,j})\leq
C_\Omega \cdot\min \{ 2^{-k}, 2^{k-2j}\}$ from \eqref{kj}, \eqref{2k},
to obtain
$$
    |I_{k,j}(\xi)|\leq \frac{C^*}{|\xi|^{3/4}}
 \cdot \Big[ \max\{  2^{k+j}, 2^{2j} \}\Big]^{3/4}
\cdot   
 \min\{ 2^{-k}, 2^{k-2j} \}\cdot {\bf 1}\big[  D(\om_\xi) \leq
  C_\Omega(2^{-k}+2^{-j}) \; \big] \; .
$$
It is easy to check by separating the $k\leq j$ and $k\ge j$
cases, that we obtain the same bound \eqref{34}
as in the first case.
Together with the trivial estimate \eqref{trivvi}, we thus have
\be
 |I_{k,j}(\xi)|\leq \frac{C^*}{ \big\langle
 |\xi|^{3/4} \big[ D(\om_\xi)\big]^{1/2}\big\rangle}
 \;
\label{34full}
\ee
in both cases.

Finally, we estimate $I_0(\xi)$. For sufficiently small $c_0$,
the boundaries of $S_0$ and $S_1$ consist of regular curves.
 We can find finitely many open balls  that
cover $S_0$ and lie within $S_1\cup S_0$. The number of the balls
is bounded by a $\Omega$-dependent number by compactness for
$a\in \cI$.
With an appropriate
partition of unity, the integral $I_0(\xi)$ is decomposed
into a finite sum of integrals of the form
$$
   \int_{D} e^{ip\cdot \xi} \;
   \psi_D(p) f(p) \rd m(p)
$$
where $D\subset S_1\cup S_0$  is a disk of radius at least $c_\Omega$
 and the smooth cutoff
function is supported on $D$. Since the Gauss curvature of $\Sigma$
is uniformly bounded from below on $D$, by standard stationary
phase estimate we obtain
\be
   |I_0(\xi)|\leq \frac{C^*}{\langle\xi\rangle} \; .
\label{I0}
\ee
Collecting  the estimates 
(\ref{tail}),   \eqref{34full}  and (\ref{I0}) 
for the decompositions \eqref{split}, \eqref{ikj},
we have proved  \eqref{fourdec}
in Theorem \ref{prop:noc}.

\medskip
\noindent
The proof of \eqref{fourdecbeta} is similar, we just sketch the
key steps. 
We define the sets $S_k$ \eqref{Sk} for all $k\ge1$
and the set $S_\infty$ will be absent. The partition of unity,
$\psi_0,\psi_1, \ldots$ consists of infinitely many functions
and $\sum_{k=0}^\infty \psi_k\equiv 1$ on the set
$\Sigma\setminus \Gamma$ of full measure.
Similarly, we extend the definition of $I_{k,j}$ \eqref{ikj}
for any $k\ge 1$ and we use the decomposition
$$
  \wh\mu(\xi) = I_0(\xi) + \sum_{k=1}^\infty \sum_{j=0}^\infty
   I_{k,j}(\xi) \; .
$$

 We now follow the previous argument.
The interpolation  \eqref{Icomb} is modified to
\be
     | I_{k,j}(\xi)| \leq \frac{C^*}{|\xi|^{\frac{3}{4}-\beta}}\cdot
\Big[ \max\{  2^{k+j}, 2^{2j} \}\Big]^{\frac{3}{4}-\beta}
\cdot     \mbox{vol} ( U_{k,j}) \; .
\label{Icombb}
\ee
First, we consider the case when
 $\mbox{vol} (U_{k,j}) \leq C_\Omega \cdot  2^{-k-j} 
D(\om_\xi)^{-1/2}$, i.e. let
$$
    \Xi= \Big\{ (k,j) \; : \; 
\mbox{vol} (U_{k,j}) \leq C_\Omega \cdot  2^{-k-j} 
D(\om_\xi)^{-1/2} \Big\}\subset \bN_+\times\bN
$$
be  the set of the corresponding indices.
For $k\leq j$, $(k,j)\in \Xi$,
we use the bound $2^{-\frac{3}{2}j} [ D(\om_\xi)]^{-1/4}$
for $\mbox{vol} ( U_{k,j})$ from \eqref{volbound}. The double summation
over $k,j$ can be performed as
$$
    \sum_{(k,j)\in\Xi \; : \; k\leq j}
 |I_{k,j}(\xi)|
  \leq \frac{C^*}{|\xi|^{\frac{3}{4}-\beta} \big[ D(\om_\xi)\big]^{1/4}}
   \sum_{k=1}^\infty \sum_{j=k}^\infty  2^{-2j\beta}
  \leq \frac{C^*\beta^{-2}}{|\xi|^{\frac{3}{4}-\beta} 
\big[ D(\om_\xi)\big]^{1/4}} \;.
$$
For $k> j$, we use  the bound
$2^{-k-\frac{3}{4}j} [ D(\om_\xi)]^{-\frac{3}{8}}$
for $\mbox{vol} ( U_{k,j})$ from the first two terms in \eqref{volbound}
and thus
$$
    \sum_{(k,j)\in\Xi \; : \; k >j}
 |I_{k,j}(\xi)|
  \leq \frac{C^*}{|\xi|^{\frac{3}{4}-\beta} \big[ D(\om_\xi)\big]^{3/8}}
   \sum_{j=0}^\infty \sum_{k=j+1}^\infty  2^{-\frac{1}{4}k}
  \leq \frac{C^*}{|\xi|^{\frac{3}{4}-\beta} 
\big[ D(\om_\xi)\big]^{3/8}} \;.
$$

On the complement of $\Xi$, 
when $D(\om_\xi)\leq C_\Omega( 2^{-k}+ 2^{-j})$, 
we again distinguish whether $k\leq j$ or $k>j$.
If $k\leq j$, then we use  $\mbox{vol} ( U_{k,j})\leq C_\Omega \cdot 2^{k-2j}$
from \eqref{kj} to obtain
\begin{align}
    \sum_{(k,j)\not\in\Xi \; : \; k \leq j}
 |I_{k,j}(\xi)|
  & \leq \frac{C^*}{|\xi|^{\frac{3}{4}-\beta} }
   \sum_{k=1}^\infty \sum_{j=k}^\infty  2^{(\frac{1}{2}-\beta)k}\cdot 
2^{-j\beta}
\cdot {\bf 1}\big[ (k,j)\not\in \Xi\big]
 \nonumber\\
&
  \leq \frac{C^*\beta^{-2}}{|\xi|^{\frac{3}{4}-\beta} 
\big[ D(\om_\xi)\big]^{\frac{1}{2}-\beta}} \;,
\nonumber
\end{align}
after replacing $2^k \leq  [D(\om_\xi)]^{-1}$ in the first
factor and using the second one, $2^{-j\beta}$, to perform 
the double summation.

Finally, if $j<k$, we use 
 $\mbox{vol} ( U_{k,j})\leq C_\Omega \cdot 2^{-k}$ from \eqref{2k}, to obtain
\begin{align}
    \sum_{(k,j)\not\in\Xi \; : \; j <k}
 |I_{k,j}(\xi)|
  & \leq \frac{C^*}{|\xi|^{\frac{3}{4}-\beta} }
   \sum_{j=0}^\infty \sum_{k=j+1}^\infty  2^{(\frac{1}{2}-\beta)j}\cdot 
2^{-k\beta}
\cdot {\bf 1}\big[ (k,j)\not\in \Xi\big]
 \nonumber\\
&
  \leq \frac{C^*\beta^{-2}}{|\xi|^{\frac{3}{4}-\beta} 
\big[ D(\om_\xi)\big]^{\frac{1}{2}-\beta}} \;.
\nonumber
\end{align}
Collecting these estimates together with \eqref{I0} and the boundedness
of $D$ and $|\wh\mu|$, we obtain
\eqref{fourdecbeta}.
$\;\;\;\Box$

\subsection{Proof of Corollary \ref{cor:L4}}

For the proof of \eqref{l4},
choose $L= \log_2 M$ in \eqref{fourdec},
 then with $\xi =r\om$, $r\ge0$, $\om\in S^2$,
\be
   \int_{|\xi|\leq M} |\wh\mu(\xi)|^4 \rd\xi  
  \leq  
  C^* + C^*L^8 \int_1^{2^L} r^2\rd r \int_{S^2} 
\frac{\rd m(\om)}{\big\langle r^3 D(\om)^2\rangle}\; ,
\label{rnu}
\ee
where $\rd m$ denotes the surface measure on $S^2$.
Using $r\leq 2^L$ we can  estimate the second integral:
\begin{align}
 \int_{S^2} 
\frac{\rd m(\om)}{\big\langle r^3 D(\om)^2\rangle} 
& \leq \frac{1}{r^3}\int_{S^2} 
\frac{ {\bf 1}\big[ D(\om)
\ge 2^{-\frac{3}{2}L}\big]}{  D(\om)^2 } \; \rd m(\om)
+  \int_{S^2} 
{\bf 1}\big[ D(\om)\leq  2^{-\frac{3}{2}L}\big]
 \; \rd m(\om)
\nonumber\\
 &\leq  \frac{1}{r^3} \sum_\pm\sum_{j=1}^{N} \int_{S^2} 
\frac{ {\bf 1}\big[ |\om\pm \nu(p^{(j)})|\ge 2^{-\frac{3}{2}L} \big]}
{ |\om\pm \nu(p^{(j)})|^2}  \; \rd m(\om) + CN 
\cdot 2^{-3L} \nonumber
\\ 
&\leq \frac{CN L}{r^3} +  CN 
\cdot 2^{-3L}
\nonumber
\end{align}
by the definition \eqref{Ddef} of $D(\om)$.
Inserting this bound into \eqref{rnu}, we obtain \eqref{l4}.

For the
 proof of \eqref{4b} we first notice that by interpolation
and $\|\wh\mu\|_\infty\leq C^*$, it is sufficient to
prove this bound for all small positive $\beta$.  
For a given $0<\beta<2/5$  we use 
 \eqref{fourdecbeta} with $5\beta/32$ instead of $\beta$ to obtain
\begin{align}
     \int_{\bR^3} |\wh\mu(\xi)|^{4+\beta}\rd\xi
    &\leq C^* + C^*\beta^{-2(4+\beta)}
    \int_{\bR^3} \frac{{\bf 1}\{ |\xi|\ge 1\} \; \rd\xi}{ \big\langle
 |\xi|^{\frac{3}{4}-\frac{5}{32}\beta}
 |D(\om_\xi)|^{\frac{1}{2}-\frac{5}{32}\beta}\big\rangle^{4+\beta}} 
\nonumber
\\
 & \leq C^* + C^*\beta^{-2(4+\beta)}
    \int_{\bR^3} 
  \frac{{\bf 1}\{ |\xi|\ge 1\} \;\rd\xi}{  |\xi|^{3+\frac{1}{16}\beta}
 |D(\om_\xi)|^{2-\frac{1}{8}\beta} + 1} 
\nonumber
\\
&
\leq C^* + C^*\beta^{-2(4+\beta)} \int_{S^2} \frac{\rd m(\om)}
{     |D(\om)|^{\frac{96-3\beta}{48+\beta}} }  \int_1^\infty
 \frac{\rd u}{u^{3+\frac{1}{16}\beta}+1}
\nonumber
\\ 
& \leq C^*\beta^{-10-2\beta} \; . \qquad \Box\nonumber
\end{align}

\section{Proof of the Four Denominator Estimate}\label{sec:fourden}
\setcounter{equation}{0}

We fix $\a$, $\eta$, $\Lambda$ and $u$ throughout the proof.
$C_\Lambda$ and $c_\Lambda$ will denote large and small
universal positive constants depending only on $\Lambda$.
We will mostly omit the $\alpha$ and $u$-dependence in the notation, all
estimates are uniform for $u\in \bR^3$ and $\a\in \bR$
 with $\tri\a\tri\ge \Lambda$.

We recall the definition of $e(p)$ from \eqref{edef}. 
The range of $e(p)$ is $[0,6]$.
Let $0\leq \chi(t)\leq 1$ be a smooth cutoff
 function on $[0,6]$ such that
$\chi(t)\equiv 1$ if $\tri t \tri \ge2\Lambda/3$
and $\chi(t)\equiv 0$ if $\tri t\tri \leq\Lambda/3$,
$|\partial^\a \chi|\leq C_\a \Lambda^{-\a}$,
where $\tri\cdot\tri$ is defined in \eqref{tripledef}.

We insert $1\equiv \chi(e(p))+ [1-\chi(e(p))]$ in the integral 
$I_{\a,\eta}(u)$ \eqref{defI}.
On the set where $1-\chi(e(p))\neq 0$ we can estimate
$$
     \frac{1}{|\a - e(p)+i\eta|}\leq C\Lambda^{-1}\; ,
$$
and once one of the denominators is eliminated, the rest
can be integrated out at the expense of $C|\log\eta|^3$, see \eqref{logeta}.

So we can focus on the term with $\chi(e(p))$.
Similarly we can insert  $\chi(e(q))\chi(e(r))\chi(e(p+q+r-u))$ as well and
we define
$$
    I=\int\frac{\chi(e(p))\chi(e(q))\chi(e(r))\chi(e(p+q+r-u))
   dpdqdr}{|\a -e(p)+i\eta||\a -e(q)+i\eta| 
|\a -e(r)+i\eta||\a -e(p-q+r-u)+i\eta|} \; .
$$
Then
\be 
    I_{\a, \eta}(u)\leq C_\Lambda|\log\eta|^3 +  I \; .
\label{Ia}
\ee

We set
\be
     I(\xi) = \int_{\Tor^3}
 \frac{e^{ip\cdot \xi}\chi(e(p))}{ |\a -e(p)+i\eta|}  \; \rd p
\label{ixi}
\ee
for $\xi\in \bR^3$, then clearly $I(\xi) = I(-\xi)$ and it is real.
Moreover
$$
   I = \frac{1}{2\pi}\int_{\bR^3} \rd\xi \; I(\xi)^4\;
 e^{-iu\cdot \xi} \leq \int_{\bR^3} \rd\xi 
 |I(\xi)|^4 \; .
$$
The function $h(p)= \chi(e(p))|\a -e(p)+i\eta|^{-1}$ in the oscillatory
integral \eqref{ixi} is
regular on scale $\eta$, 
$$
   |\partial^\beta h(p) |\leq \frac{C_{\Lambda, \beta}\cdot
\eta^{-|\beta|} }{
   |\a-e(p)+i\eta|}
$$
for any multiindex $\beta$.
Thus, by a standard stationary phase estimate and \eqref{logeta},
 we easily see that
$$
      |I(\xi)|\leq \frac{C_\Lambda|\log\eta|}{ \langle \eta|\xi|\rangle} \; ,
$$
therefore
\be
     I\leq C_\Lambda|\log\eta|^4+\int_{|\xi|\leq \eta^{-4}}
 |I(\xi)|^4\rd\xi \; .
\label{Iaa}
\ee

By the coarea formula
$$
    I(\xi) = \int_0^6 \frac{\chi(a)\rd a}{|\a-a+i\eta|} \; \wh\mu_a(\xi) 
$$
with
$$
    \wh\mu_a(\xi)= \int_{\Sigma_a} \frac{e^{ip\cdot\xi}}{|\nabla e(p)|}\;
 \rd m_a(p)\; ,
$$
where we recall that
 $\rd m_a$ is the uniform surface measure on the set $\Sigma_a=\{ p \; : \;
e(p)=a\}\subset \Tor^3$. 
Clearly $\wh\mu_a(\xi)$ is an integral of the form
\eqref{whmu} with  $f(p)=|\nabla e(p)|^{-1}$.
 Note that
\be
      \tri a\tri^{1/2} \leq C\Big(\sum_{j=1}^3
    \sin^2 p_j  \Big)^{1/2}= C |\nabla e(p)| \; .
\label{simm}
\ee
Thus for $\tri a\tri\ge\Lambda/3$,
the function $|\nabla e(p)|$ on the set $\Sigma_a$ is separated
away from zero and is smooth with derivatives bounded 
uniformly in $a$ (depending only on $\Lambda$), so $|\nabla e(p)|^{-1}$
is smooth.

The main technical result is the following special case of 
Corollary \ref{cor:L4} for the family of level sets
 $\{ e(p)=a\} $ with values in the compact
set $\cI= \{ a\in [0,6] \; : \; \tri a\tri \ge
\Lambda/3\}$.

\begin{proposition}\label{prop:J} Let $0 <\Lambda<1/2$.
For any $a$ with $\tri a\tri \ge \Lambda$, we have
$$
    \int_{|\xi|\leq \eta^{-4}} |\wh\mu_a(\xi)|^4 
\rd\xi \leq C_\Lambda|\log\eta|^{10} \; .
$$
\end{proposition}

The proof amounts to checking the
assumptions in Corollary \ref{cor:L4}. Assumption 1 (formula \eqref{ass1})
has been checked in \eqref{simm}.
The other three assumptions will be proven
 starting from the next section.

{F}rom this  Proposition  and \eqref{Ia}, \eqref{Iaa},
  the Four Denominator Estimate
\eqref{4den} easily follows. By Jensen's inequality,
\begin{align}
    \int_{|\xi|\leq \eta^{-4}} \rd\xi |I(\xi)|^4
  & = \int_{|\xi|\leq \eta^{-4}} \rd\xi \Bigg| 
 \int_0^6 \frac{\chi(a)\rd a}{|\a-a+i\eta|} \;  \wh\mu_a(\xi)
   \Bigg|^4  \nonumber
\\
  &\leq 
  \Bigg(  \int_0^6 \frac{\chi(a)\rd a}{|\a-a+i\eta|}\Bigg)^3
   \int_{|\xi|\leq \eta^{-4}}\rd\xi   
\int_0^6 \frac{\chi(a)\rd a}{|\a-a+i\eta|}  \; | \wh\mu_a(\xi)|^4
  \nonumber\\
& \leq C_\Lambda |\log\eta|^{14} \; 
\end{align}
by applying Proposition \ref{prop:J} with $\Lambda/3$ instead of $\Lambda$
and by recalling the support of $\chi$. $\;\;\Box$

\subsection{The geometry of the isoenergy surface}

We use the notation  $p=(p_1, p_2, p_3)\in\Tor^3$ and
$$
 s_j = \sin p_j, \qquad c_j =\cos p_j \; .
$$
We work on the surface $\Sigma_a$ given by
\be
    e(p)= 3-(c_1+ c_2 + c_3) =a \; , \qquad p\in \Tor^3
\label{surf}\ee
and we assume that $\tri a\tri \ge \Lambda$.  
Let $K(p)$ be the Gauss curvature and $H(p)$ 
be the mean curvature of the surface $\Sigma_a$ at the  point $p\in \Sigma_a$.
The following Lemma is proved in Section \ref{sec:tech}.

\begin{lemma}\label{lemma:curvs}
The Gauss curvature $K$ of $\Sigma_a$ is given by 
\be
    K =  \frac{ s_1^2 c_2c_3 + s_2^2 c_1 c_3 + s_3^2 c_1c_2}{(s_1^2+
        s_2^2+s_3^2)^2}
\label{Kcur}\ee
and the mean curvature is
\be
    H=\frac{1}{\sqrt{s_1^2+s_2^2 +s_3^2}}\Big( 3-a - \frac{s_1^2c_1
        + s_2^2c_2 + s_3^2 c_3}{s_1^2+s_2^2 +s_3^2} \Big)
\label{Hcur}
\ee
For $a\in (0,2)\cup (4,6)$  the Gauss
curvature satisfies
\be
     K\ge C\tri a \tri^2
\label{Klow}
\ee
with some universal constant,
in particular $\Sigma_a$ is uniformly convex.
The surface $\Sigma_a$ has a flat umbilic point
if and only if $a=3$.
\end{lemma}

The following lemma lists some properties of the normal map,
 $\nu : \Sigma_a\to S^2$, given by
$$
     \nu(p)= \frac{\nabla e(p)}{|\nabla e(p)|} \; .
$$
In particular it verifies Assumption 3 (formula \eqref{ass3}).
The proof is given in Section \ref{sec:tech}.

\begin{lemma}\label{lemma:mult}
The map $\nu(p)$ is surjective. 
It is also bijective for $a\in (0,2)$ or $a\in (4,6)$.
For $a\in (2,4)$, the set of preimages 
 $\{ p\; : \; \nu(p) =\nu\}$ have cardinality  at most 64
for any $\nu\in S^2$.
The derivative of the (local) inverse map, $p'(\nu)$,
is bounded from above 
\be
  \| p'(\nu)\|\leq \frac{C}{|K(p(\nu))| \cdot\tri a \tri} \;.
\label{jacb1}
\ee
%(see \eqref{jacb} in the general case).
\end{lemma}

\bigskip
\noindent
The following Proposition estimates the uniformly convex case.
\begin{proposition}
Let $a\in [\Lambda,2-\Lambda]\cup [4+\Lambda, 6-\Lambda]$,
then
\be
    |\wh\mu_a(\xi)|\leq 
 \frac{C_\Lambda}{ \langle\xi\rangle} \; .
\label{J}
\ee
\end{proposition}
This proposition is standard in harmonic analysis, see
e.g. Theorem 1. Section VIII.3.1 of \cite{St}. The uniformity 
of the constant in $a$ follows from the uniform bound
(\ref{Klow}) on the curvature and
from the uniform bounds on the derivatives
of $|\nabla e(p)|^{-1}$.

\bigskip
\noindent

{F}rom now on we work with the  $a\in [2+\Lambda, 3-\Lambda]\cup
[3+\Lambda, 4-\Lambda]$ case. 
The next lemma verifies Assumption 2 (formula \eqref{ass2})
and is proven in Section \ref{sec:tech}.

\begin{lemma}\label{lemma:wedge} There exists a  positive constant
$c^*_\Lambda\ll 1$ such that whenever $K(p)=0$,
%\leq c^*_\Lambda$, 
then
\be
     |\nabla e(p)\times \nabla K(p)|\ge c^*_\Lambda \; .
\label{eK}
\ee
\end{lemma}

\bigskip
\noindent
Recall that
at every point $p\in \Sigma_a$ we defined the projection
$P=P(p)= I-|\nu\rangle\langle \nu|$ from
 $T_p \Tor^3$ onto  the subspace orthogonal to the normal
vector $\nu=\nu(p)$ that can be identified with  $T_p\Sigma_a$.
Let $A=A(p)=e''(p)$ be the Hessian matrix, it
is diagonal with entries $c_1, c_2, c_3$.

Introduce the notation
\be
    M= K|\nabla e|^4 = s_1^2 c_2c_3 + s_2^2 c_1c_3 + s_3^2c_1c_2\; .
\label{Mdef}
\ee
The  unit tangent vector of $\Gamma$ is given by 
$$
     w=w(p)=  \frac{\nabla e(p) 
\times \nabla M(p)}{|\nabla e(p) \times \nabla M(p)|} \; .
$$
Note that this definition slightly differs from 
\eqref{wdef}, but it actually defines the same vectorfield
on $\Gamma$ since $\nabla e\neq 0$. The following Lemma verifies Assumption 4*,
or, equivalently, Assumption 4 (see formulae  \eqref{ass4*}
and \eqref{ass4}).

\begin{lemma}\label{lemma:tr}
There exist positive constants $c_\Lambda$, $C_\Lambda$
such that for any $a\in [2,4]$, $\tri a\tri\ge \Lambda$,  there exist
 $1\leq N_a\leq C_\Lambda$ tangential
points, $p^{(1)}, p^{(2)}, \ldots , p^{(N_a)}$ 
on the curve $\Gamma_a$ such 
that   $ PAP w(p^{(j)})=0$ and
\be
    \| PAP w(p)\|\ge c_\Lambda\cdot \min\{ |p-p^{(j)}|\; : \; j=1, 2,\ldots N_a
    \} \;, \qquad p\in \Gamma_a \; .
\label{PAPlo}
\ee
\end{lemma}

\bigskip
\noindent

{\it Proof of Lemma \ref{lemma:tr}.} 
 Define the unit vector $\mu=(\mu_1, \mu_2, \mu_3)$
by its components 
$$
   \mu_j = \tau(p)\tan p_j,\qquad
  \tau(p)=\frac{1}{\sqrt{\tan^2 p_1+ \tan^2 p_2+\tan^2p_3}}
$$
on $\Gamma$ away from the $p_j=\pm\frac{\pi}{2}$ hyperplanes.
Since on $\Gamma$
\be
     0= s_1^2c_2c_3+s_2^2c_1c_3+ s_3^2c_1c_2,
\label{ko}
\ee
if $p_1\to \pm\frac{\pi}{2}$, i.e.
 $c_1\to 0$, then either $c_2$ or $c_3$ must go to zero as well.
Assume that $c_2\to 0$, i.e. $p_2\to \pm\frac{\pi}{2}$ as
well. Since $c_1+c_2+c_3=3-a \neq 0$, $c_3$
remains separated away from zero in the neighborhood
$c_1, c_2\sim0$. From \eqref{ko}
\be
   0=(c_1+c_2)c_3 +s_3^2c_1c_2 - c_1c_2c_3(c_1+c_2) \; ,
\label{cc}
\ee
thus $(c_1+c_2)/c_2\to0$ as $c_1, c_2\to 0$.
Therefore
\be
    \frac{\mu_2}{\mu_1} = \frac{s_2}{s_1}\Big( \frac{c_1+c_2}{c_2} -1\Big)
    \to -1
\label{mumu}
\ee
in the neighborhood $p_1, p_2 \sim \frac{\pi}{2}$ and similar
relations hold at the other three points where $c_1, c_2\sim 0$.
Since $\tau\to\infty$ and $\mu_3\to 0$, the relation \eqref{mumu}
shows that $\mu$ extends continuously to the points where $c_1=0$.
Similar relation holds for the other points where $\mu$ has
a virtual singularity, thus $\mu$ is actually a continuous unit 
vectorfield on $\Gamma$.

Straightforward calculations give the following relations
on the curve $\Gamma$
$$
        \mu \perp \nu, \qquad \nu\perp w,\qquad
        A\mu \; \Vert\; \nu\; .
$$
In particular, $PAP\mu=0$ since $P\mu=\mu$ and $P\nu=0$,
so $\mu$ is the kernel direction of the Gauss map.
Let $\wt\mu$ be the unit vector orthogonal to both $\nu$ and
$\mu$, i.e. $\wt\mu$ is the direction of the other principal
curvature.

{F}rom (\ref{eK}) it follows that $K(p)$ has only
a single zero on  $\Gamma$, i.e. only one of the
principal curvatures is zero. The other principal
curvature therefore is bounded from below by $c_\Lambda$
using the compactness of the domain 
$D_\Lambda=\{ p\in \Tor^3\; : \;  \tri e(p)\tri\ge \Lambda\}$:
$$
    \| PAP \wt\mu\|\ge c_\Lambda\; .
$$
Decomposing $w= (w\cdot \mu)\mu + (w\cdot \wt\mu)\wt\mu$, we get
$$
    \| PAP w\|\ge c_\Lambda |w\cdot \wt\mu| = c_\Lambda|w\times \mu|\; .
$$
By using the definition of $w$, the boundedness of $|\nabla e\times \nabla M|$
and $\nu\perp\mu$, we have
$$
    |w\times \mu|\ge c_\Lambda |(\nu\times \nabla M)\times \mu|
   = c_\Lambda |\mu\cdot \nabla M| \; .
$$

Therefore we have to prove that $\mu$ can be orthogonal to $\nabla M$
only at finitely many points on $\Gamma_a$ and the angle between them
  changes at least linearly as we move away from these points.

Let $\delta\ll 1$ be a sufficiently small positive number depending
only on $\Lambda$.
If $|c_1c_2c_3|\leq \delta^6$, then
at least one of the $c_j$'s is smaller than $\delta^2$, 
say $|c_1|\leq \delta^2$. In
this case  $|K|\ge c_\Lambda |c_2c_3|-C_\Lambda\delta^2$ by using \eqref{Kcur}.
On the set  $|K|\leq \delta^2$ it follows that either $|c_2|\leq 
C_\Lambda\delta$
or $|c_3|\leq C_\Lambda\delta$. Suppose $|c_2|\leq C_\Lambda\delta$, then
$|c_3-(3-a)|=|c_1+c_2|\leq C_\Lambda\delta$.
By permuting the indices we obtain that away from a $C_\Lambda\delta$
neighborhood of the set
$$
E_a=\Big\{ (0,0,3-a), (0,3-a, 0), (3-a,0,0)\Big\}
$$
we have $|c_1c_2c_3|\ge \delta^6$. Therefore 
we distinguish two cases:

\bigskip
\noindent
{\it Case 1: $(c_1, c_2, c_3)$ is in a $C_\Lambda\delta$
 neighborhood of $E_a$. }

\bigskip
\noindent
{\it Case 2: $|c_1c_2c_3|\ge \delta^6$ }

\bigskip
\noindent
Now we analyze these cases separately.

\bigskip
\noindent
{\it Case 1.} The points in $E_a$ correspond to
 vectors $p^*=(p_1^*,p_2^*,p_3^*)$
where two components are $\pm\pi/2$ and one component is $\pm\cos^{-1}(3-a)$.
Let $p^*$ be one of these finitely many points,
and we will study a small neighborhood of $p^*$.
For definiteness, let $c_1=c_2=0$ at $p^*$.

 We need to compute the variation of $|\mu\cdot\nabla M|$
along the curve $K=0$ near this point. 
At an arbitrary point $p\in \Gamma$ near $p^*$ we have
\be
    \mu\cdot \nabla M=\tau\Bigg[\frac{s_1^2}{c_1}(2c_1c_2c_3 - s_2^2c_3
  - s_3^2c_2) +\frac{s_2^2}{c_2}(2c_1c_2c_3 - s_1^2c_3
  - s_3^2c_1) +\frac{s_3^2}{c_3}(2c_1c_2c_3 - s_2^2c_1
  - s_1^2c_2) \Bigg] \; .
\label{tauu}
\ee
Set $\e=\sqrt{c_1^2+c_2^2}\ll 1$, it is clear that $|p-p^*|\sim \e$.
An explicit calculation shows that
$$
    \tau =\frac{|c_1c_2|}{\sqrt{c_1^2+c_2^2}} (1+ O(\e^2))
$$
and
$$
    \big|\, (\mu\cdot\nabla M)(p)\,\big|=
\frac{|c_1^2+c_2^2-c_1c_2|}{\sqrt{c_1^2+c_2^2}} |s_3|^2(1+ O(\e^2)) \; .
$$
Thus $\mu\cdot \nabla M \to 0$ as $p\to p^*$, but $\mu$ and $\nabla M$
are regular, thus $\mu\cdot \nabla M$ vanishes at $p^*$,
so $p^*$ is a tangential point.
In its small neighborhood,
$$
 \big|\, (\mu\cdot\nabla M)(p)\,\big|\ge c_\Lambda \e \ge c_\Lambda |p-p^*|
$$
by using $c_1^2+c_2^2 - c_1c_2 \ge \frac{1}{2}(c_1^2+c_2^2)$
and that $s_3^2\ge 1-(3-a)^2\ge 1-\Lambda^2$.
 Therefore we can add
these finitely many points $p^*$ to the collection tangential points,
and (\ref{PAPlo}) will hold in a small, $\Lambda$-dependent 
 neighborhood of $p^*$.

\bigskip
\noindent

{\it Case 2.} 
In this case   $|c_j|\ge\delta^2$ for each $j$, so we have
 $\tau\ge c_\Lambda\delta^2$,
 so it is sufficient to
give a lower bound on $\tau^{-1}|\mu\cdot\nabla M|$.
We use the formula (\ref{tauu}).
On $\Gamma$ we have
$$
     \frac{s_1^2}{c_1} +\frac{s_2^2}{c_2} + \frac{s_3^2}{c_3} =0
$$
from \eqref{Kcur}, so
\be
     \frac{1}{c_1}+\frac{1}{c_2}+\frac{1}{c_3}=c_1+c_2+c_3=3-a \; .
\label{reci}
\ee
This is actually the equation of $\Gamma=\{ K=0\}\cap\Sigma_a$.
Thus we rewrite 
$$
    2c_1c_2c_3  - s_2^2c_3
  - s_3^2c_2= 2((3-a)-c_2-c_3)c_2c_3 - (1-c_2^2)c_3-(1-c_3^2)c_2
$$
$$
   = c_2c_3\Big[ 2(3-a)- c_2-c_3-\frac{1}{c_2}-\frac{1}{c_3}\Big] 
   = c_2c_3\Big(\frac{1}{c_1}+c_1\Big)
$$
by using \eqref{reci}, and similarly for the other two terms
in \eqref{tauu}.
Therefore
\be
   \tau^{-1}\mu\cdot\nabla M
   = \frac{(1-c_1^4)c_2c_3}{c_1^2} + \frac{(1-c_2^4)c_1c_3}{c_2^2}
+\frac{(1-c_3^4)c_2c_1}{c_3^2}\; .
\label{1*}
\ee

First we consider the possible solutions to the equations (\ref{reci})
and 
\be
  0=\mu\cdot\nabla M= \frac{\tau}{(c_1c_2c_3)^2}
\Big[ (1-c_1^4)c_2^3c_3^3 + (1-c_2^4)c_1^3c_3^3+(1-c_3^4)c_2^3c_1^3\Big]
\label{mum}
\ee
 Viewing $c_1, c_2, c_3$ as
three independent variables, we compute the Jacobian  of the map
$$
   \Phi(c_1,c_2,c_3)= \begin{pmatrix} 
   c_1+c_2+c_3\cr\cr
   c_1^{-1}+c_2^{-1}+c_3^{-1}\cr\cr
   (1-c_1^4)c_2^3c_3^3 + (1-c_2^4)c_1^3c_3^3+(1-c_3^4)c_2^3c_1^3
\end{pmatrix}
$$
defined away from $\{ c_1=0\}\cup \{c_2=0\}\cup \{c_3=0\}$.
We use that
$$
    \frac{\partial}{\partial c_1}\Big[
(1-c_1^4)c_2^3c_3^3 + (1-c_2^4)c_1^3c_3^3+(1-c_3^4)c_2^3c_1^3\Big]
 = -4c_1^3c_2^3c_3^3 +3c_1^2\Big[ c_3^2(1-c_2^4) + c_2^2(1-c_3^4)\Big]
$$
$$
   =-4c_1^3c_2^3c_3^3 -\frac{(1-c_1^4)c_2^3c_3^3}{c_1}
$$
on the solution set $\mu\cdot \nabla M=0$.
After a somewhat tedious calculation we obtain for the Jacobi determinant
$$
     \Big|\frac{\partial\Phi}{\partial c}\Big|
   =\Bigg| \mbox{det}\begin{pmatrix}
   1 & 1 & 1\cr\cr
     c_1^{-2}&c_2^{-2} &c_3^{-2}\cr\cr
   \frac{(1-c_1^4)c_2^3c_3^3}{c_1} &
  \frac{(1-c_2^4)c_1^3c_3^3}{c_2} &\frac{(1-c_3^4)c_2^3c_1^3}{c_3} 
\end{pmatrix} \Bigg|=
 \frac{|c_1^2-c_2^2||c_2^2-c_3^2||c_3^2-c_1^2|}{|c_1c_2c_3|} \; 
$$
whenever $\mu\cdot \nabla M=0$.
\begin{lemma}\label{lemma:jacno}
For $\tri a\tri \neq 0$,
the Jacobian $|\partial\Phi/\partial c|$ does not vanish
on the solution set
$\Phi(c_1, c_2, c_3)=( 3-a, 3-a, 0)$.
\end{lemma}

{\it Proof.} Suppose, on the contrary, that the Jacobian is zero, say
$c_1^2=c_2^2$. If $c_1=-c_2$, then $c_3=3-a=c_3^{-1}$
from \eqref{reci} i.e.
$c_3=\pm 1$, so $\tri a\tri=0$. If $c_1=c_2$, then  from \eqref{reci}
\be
2c_1+c_3=2c_1^{-1}+
c_3^{-1}=3-a,
\label{3-a}
\ee
 moreover, from \eqref{1*},
$$
    2(1-c_1^4)c_1^3c_3^3 + (1-c_3^4)c_1^6=0 \; .
$$
From this and $2c_1+c_3=3-a$ we obtain
$$
    2c_1^{-3}+
c_3^{-3}=3-a \; .
$$
Combining this with (\ref{3-a}) we get $(2c_1^{-1}+
c_3^{-1})^2=(2c_1+c_3)(2c_1^{-3}+
c_3^{-3})$ thus $c_1^2=c_3^2$. If $c_1=c_3$, then $3c_1=3-a=3c_1^{-1}$,
i.e. $c_1^2=c_2^2=c_3^2=1$ and $\tri a\tri =0$. If $c_1=-c_3$,
then we have $c_2=3-a=c_2^{-1}$ and again $\tri a\tri =0$.
This completes the proof of Lemma \ref{lemma:jacno} $\;\;\Box$

Using this lemma and 
the compactness of $c_j\in [-1,-\delta^2]\cup[\delta^2, 1]$,
 $a\in [2+\Lambda, 4-\Lambda]$,
we obtain that the Jacobian $|\partial\Phi/\partial c|$
is always bounded from below
by a positive $\Lambda$-dependent constant on the solution set
$\Phi(c_1, c_2, c_3)=( 3-a, 3-a, 0)$, uniformly in $\tri a\tri \ge \Lambda$.
 Then by the inverse function theorem
and compactness 
we obtain that the solution set consists of finitely many  disjoint
branches $\{ p^{(1)}(a), 
p^{(2)}(a), \ldots p^{(N)}(a)\}$. Moreover, by using
the relation between $\mu\cdot \nabla M$ and the
third component of $\Phi$ (see (\ref{mum})), and
the fact that on $\Gamma$ the first two components
are constant $3-a$, 
the bound (\ref{PAPlo}) holds with a sufficiently small
$c_\Lambda$.
This completes the proof of Lemma \ref{lemma:tr} 
$\;\;\Box$

\subsection{Proof of the technical lemmas}\label{sec:tech}

{\it Proof of Lemma \ref{lemma:log}.} By the coarea formula
$$
   \int_{\Tor^3}\frac{\rd p}{|\a - e(p) + i\eta|} 
  = \int_0^6 \frac{\Phi(a)\rd a}{|\a - a + i\eta|} \; , 
\quad \mbox{with} \quad \Phi(a)= \int_{\Sigma_a} \frac{\rd m_a}{|\nabla e|}\;.
$$
The estimate \eqref{logeta} will follow from the boundedness of $\Phi(a)$.
Away from the critical points of $e(p)$, $|\nabla e(p)|$ is 
separated away from zero, thus $\Phi$ is bounded.
 There are eight critical points, 
each $p_j$ can be either 0 or $\pi$ (recall that $\pi=-\pi$ on the torus).
Two of them are elliptic, six are hyperbolic. With a regular bijection,
a small neighborhood of the
critical points on the surface $\Sigma_a$ can be brought into
a normal form $f(x)=x_1^2+x_2^2+x_3^2=\e$ or $f(x)=x_1^2+x_2^2-x_3^2=\e$
with $|\e|\ll 1$, $|x|\ll 1$.  Explicit calculation 
shows that in both cases
$$
       \int_{f=\e} \frac{{\bf 1}[|x|\leq \delta] \; \rd m_\e(x)}{|\nabla f(x)|}
$$
is uniformly bounded as $\e, \delta\to 0$. Here $\rd m_\e$
denotes the surface measure on the level set $\{ x\; : \; f(x)=\e\}$.
 $\;\;\Box$

\medskip
\noindent
{\it Proof of Lemma \ref{lemma:curvs}.} 
Since $|\nabla e(p)|\ge C\tri a \tri^{1/2} \ge C_\Lambda$, we can 
express one of the three variables in terms of the other two
in  a local chart. We work on a chart where $\Sigma_a$ is
given as a function $p_3=p_3(p_1, p_2)$.

It is well known that the Gauss curvature
of a surface given locally by a function $z=f(x,y)$ is
\be
   K = (1+|\nabla f|^2)^{-2} \mbox{det} f''\; .
\label{curv}\ee
while the mean curvature is given by
\be 
   H = \mbox{div} \Big(\frac {\nabla f}{\sqrt{1+|\nabla f|^2}}\Big)\; .
\label{Hcurv1}
\ee

 Differentiate  (\ref{surf}) with respect to $p_1$:
\be
  s_3\frac{\partial p_3}{\partial p_1} + s_1 =0 \; .
\label{1der}\ee
The second $p_1$ derivative gives
$$
  s_3\frac{\partial^2 p_3}{\partial p_1^2} 
  +c_3\Big(\frac{\partial p_3}{\partial p_1}\Big)^2
+ c_1 =0\; ,
$$
so
$$
   \frac{\partial^2 p_3}{\partial p_1^2} =
   -\frac{s_2^2 c_3 +s_3^2 c_2}{s_3^3}\; ,
$$
and similarly
$$
   \frac{\partial^2 p_3}{\partial p_2^2} =
   -\frac{s_1^2 c_3 +s_3^2 c_1}{s_3^3} \; .
$$
For the mixed derivative, the $p_2$ derivative of (\ref{1der}) gives
$$
  s_3 \frac{\partial^2 p_3}{\partial p_2\partial p_1}
 +c_3\frac{\partial p_3}{\partial p_1}\cdot
\frac{\partial p_3}{\partial p_2}=0 \; ,
$$
therefore
$$
 \frac{\partial^2 p_3}{\partial p_2\partial p_1} =-\frac{s_1s_2c_3}{s_3^3}\;.
$$
Collecting all these information, one obtains (\ref{Kcur}) 
and (\ref{Hcur})
from (\ref{curv}) and \eqref{Hcurv1}.

For the convexity, can assume that $a\in (0 ,2)$, the other case
follows by symmetry. Then
$$
   s_1^2 c_2c_3 + s_2^2 c_1 c_3 + s_3^2 c_1c_2 
  = (1-c_1)(1-c_2)(1-c_3) + (2-a)(1-c_1c_2c_3) 
$$
Since $c_1+c_2+c_3=3-a\in (1+\tri a\tri, 3-\tri a\tri)$,
 at least two of the $c_j$'s
must be nonnegative. If all of them are nonnegative, then
$c_1c_2c_3\leq [(c_1+c_2+c_3)/3]^3 \leq 1-C\tri a \tri $,
otherwise $c_1c_2c_3\leq 0$. In both cases we obtain
$$ 
 s_1^2 c_2c_3 + s_2^2 c_1 c_3 + s_3^2 c_1c_2 \ge C\tri a \tri^2
$$
with some universal constant.
The uniform convexity follows from the lower bound on $K$
and the uniform upper bound 
\be
   |H|\leq C\tri a \tri^{-1/2}
\label{Hb}
\ee
on the mean curvature (see \eqref{simm} and \eqref{Hcur}),
 since, if $\kappa_1, \kappa_2$
are the two curvatures, then
$$ 
    \kappa_1+\kappa_2\leq C\tri a \tri^{-1/2}, \qquad
   \kappa_1\kappa_2\ge C\tri a \tri^2
$$
imply $\kappa_i\ge C\tri a \tri^{-5/2}$.

Finally, for the statement on the flat umbilic points, we
set $\lambda=3-a$ and it is sufficient
to consider $|\lambda|<1$.
 $\Sigma_a$ has a flat umbilic point at $p$ if and only
if $H=K=0$.
Based upon \eqref{Kcur} and \eqref{Hcur},
 in terms of $c_1, c_2, c_3$ it  means that
\begin{align}
  c_1+c_2 + c_3 & =  \lambda \nonumber\\
 c_1c_2 + c_2c_3 + c_3c_1 & =  \frac{2\lambda^2}{\lambda^2-3} \nonumber\\
 c_1c_2 c_3 & = \frac{2\lambda^2}{\lambda^2-3} \;. \nonumber
\end{align}
In other words, $c_1, c_2, c_3$ are solutions of
the cubic equation $f(c) = c^3-\lambda c^2 + \frac{2\lambda^2}{\lambda^2-3}c
- \frac{2\lambda^2}{\lambda^2-3}$. It is a straighforward algebraic
exercise to check that the discriminant of this equation 
is positive unless $\lambda=0$, hence 
it cannot have three real roots. If $\lambda=0$, $a=3$,
then $c_1=c_2=c_3=0$ and the eight 
points $(\pm\frac{\pi}{2}, \pm\frac{\pi}{2}
\pm\frac{\pi}{2})$ are indeed flat umbilic points.
$\;\;\Box$

\bigskip
\noindent
{\it Proof  of Lemma \ref{lemma:mult}.}  When the level sets are convex
($a\in (0,2)$ or $a\in (4,6)$), the bijectivity follows directly
from geometry (the proof below can be also modified to see this).
Otherwise, for the bijectivity,  we have to show
that the equations
\be
     \nu_j = \frac{s_j}{\sqrt{s_1^2 + s_2^2 + s_3^2}}, \qquad j=1,2,3\; ,
\label{nu}\ee
$$
    c_1 + c_2 + c_3 = 3-a 
$$
have a solution for any given $\nu\in S^2$ and $a\in (2,4)$. Let $\lambda=
1/\sqrt{s_1^2 + s_2^2 + s_3^2}$, then the constraint equation means that
\be
    f_{\pm\pm\pm}(\lambda)=\pm \sqrt{ 1- \Big( \frac{\nu_1}{\lambda}\Big)^2}
    \pm \sqrt{ 1- \Big( \frac{\nu_2}{\lambda}\Big)^2}
   \pm \sqrt{ 1- \Big( \frac{\nu_3}{\lambda}\Big)^2}=3-a \; .
\label{constr}
\ee
The three signs can be chosen independently.
By symmetries, we can assume that $\nu_j\ge 0$ and
$\nu_1 \geq \nu_2\geq\nu_3$. We can also assume that $3-a \in [0,1)$
(by symmetry), and we will choose the signs as follows:
$$
   f_{++-}(\lambda)= \sqrt{ 1- \Big( \frac{\nu_1}{\lambda}\Big)^2}
    +\sqrt{ 1- \Big( \frac{\nu_2}{\lambda}\Big)^2}
   - \sqrt{ 1- \Big( \frac{\nu_3}{\lambda}\Big)^2}=3-a
$$
We first solve this equation for $\lambda$. If $\lambda=\nu_1$, then
we have
$$
  f_{++-}(\nu_1)= \sqrt{ 1- \Big( \frac{\nu_2}{\nu_1}\Big)^2}
   - \sqrt{ 1- \Big( \frac{\nu_3}{\nu_1}\Big)^2}\leq 0
$$
As $\lambda\to \infty$, we have
$$
  \lim_{\lambda\to\infty} f_{++-}(\lambda)=1
$$
therefore, by continuity, the equation $f_{++-}(\lambda)=3-a$ has a solution.
With this $\lambda\in [\nu_1,\infty)$, we can find $p_j\ge 0$ such that
\be
   \sin p_j = \frac{\nu_j}{\lambda}
\label{sinnu}
\ee
and the sign of $\cos p_j$ is the one given by the sign choices in $f$,
therefore $p\in \Sigma_a$.  This shows the surjectivity of the normal map
$\nu(p)$ for each choice of the signs.

Now we show that (\ref{constr}) has  at most 8 solutions for $\lambda$.
Bringing one of the square roots onto the
right side and squaring this equation, we obtain a relation
that contains two square roots. With two more squarings, we
obtain a polynomial of degree eight in $\lambda^{-1}$, therefore
the number of solutions is at most 8 for each sign combinations.
For each
 each solution $\lambda$, the equations \eqref{sinnu} have a
unique solution, given the sign choice of $\cos p_j$.
This gives at most 64 preimages of the normal map.

%Finally, the Jacobian is straight-forward local calculation
% from (\ref{nu}), but it also follows from
% general facts in differential geometry.
For the bound (\ref{jacb1}) we first notice 
from (\ref{Kcur}) and (\ref{Hcur}) that $|K|, |H|\leq C\tri a\tri^{-1}$,
therefore $|\kappa_1|, |\kappa_2|\leq C\tri a\tri^{-1}$
holds as well for the two principal curvatures.
Then
$$
     \| p'(\nu)\| =\max |\kappa_j|^{-1} \leq \frac{|\kappa_1|+ 
|\kappa_2|}{|\kappa_1||\kappa_2|} \leq \frac{C}{|K|\cdot \tri a \tri} 
\; . \qquad
 \Box
$$

\bigskip{\it Proof of Lemma \ref{lemma:wedge}.}
Recalling the definition of $M$ (\ref{Mdef}), we have
$$
   \nabla K = |\nabla e|^{-4}\nabla M - 4M|\nabla e|^{-5} \nabla |\nabla e|
$$
and
\be
 |\nabla e(p)\times \nabla K(p)|\ge|\nabla e|^{-4}
 \Bigg( |\nabla e(p)\times \nabla M(p)| -  C|K|\Bigg)
\label{EK}
\ee
with a universal constant, using that $ \nabla |\nabla e|$ and $|\nabla e|$
are uniformly bounded.

We compute 
$$
  \nabla M(p) = \begin{pmatrix}  s_1(2c_1c_2c_3 - s_2^2 c_3 - s_3^2 c_2) \cr
   s_2(2c_1c_2c_3 - s_1^2 c_3 - s_3^2 c_1) \cr
  s_3(2c_1c_2c_3 - s_2^2 c_1 - s_1^2 c_2) \; .
 \end{pmatrix}
$$
Simple calculation shows that on the surface $c_1+c_2+c_3=3-a$ we have
\be
      \nabla e(p) \times \nabla M(p) 
     = \begin{pmatrix} s_2s_3(c_2-c_3)(1-(3-a)c_1) \cr
 s_1s_3(c_3-c_1)(1-(3-a)c_2) \cr s_1s_2(c_1-c_2) (1-(3-a)c_3) \end{pmatrix}\;,
\label{def:w} 
\ee
therefore
\be
   |\nabla e(p) \times \nabla M(p)|\ge \tri a\tri\Big[ |s_2s_3(c_2-c_3) |
+  |s_1s_3(c_3-c_1)|+| s_1s_2(c_1-c_2)| \Big]
\label{Mwed}
\ee
using $|1-(3-a)c_j|\ge 1 - |3-a|\ge \tri a\tri$ for $a\in (2,4)$.

\begin{lemma}\label{lemma:U} There exists a positive universal constant 
$c_\Lambda$
such that
$$
U=\Big[ |s_2s_3(c_2-c_3) |
+  |s_1s_3(c_3-c_1)|+| s_1s_2(c_1-c_2)| \Big]\ge c_\Lambda\; ,
$$
whenever $|M|\leq c_\Lambda$, $c_1+c_2+c_3=3-a$ and $\tri a\tri\ge \Lambda$.
\end{lemma}

From \eqref{EK} and \eqref{Mwed} we  thus obtain
\be
 |\nabla e(p) \times \nabla K(p)|\ge \tri a\tri c_\Lambda =:c'_\Lambda>0
\label{Ktran}
\ee
on the zero curvature line $K(p)=0$. $\;\;\Box$.

\bigskip
\noindent
{\it Proof of Lemma \ref{lemma:U}.}
We will show that $U$ and $M$ never vanish at the same point.
Since these functions are continuous on the compact
domain $D_\Lambda=\{ p\in \Tor^3\; : \;  \tri e(p)\tri\ge \Lambda\}$,
we obtain that $c_\Lambda=\frac{1}{2}\inf_{D_\Lambda} |M|+|U|>0$.

Suppose that $U=0$. If $c_1=c_2=c_3$, then from $M=0$
and $s_1^2+s_2^2+s_3^2 >0$ it follows that $c_1=c_2=c_3=0$,
but then $a= 3$, $\tri a\tri =0$.

If two of the three $c_1, c_2, c_3$ coincide, then we can
assume by symmetry that $c_1=c_2\neq c_3$ and then $s_2s_3=s_1s_3=0$
from $U=0$.
Therefore either $s_3=0$ or $s_1=s_2=0$. In the first case
it follows from $M=0$ and $s_1^2+s_2^2>0$ that $c_1=c_2=0$,
but then $c_1+c_2+c_3= \pm 1$, so $\tri a\tri=0$.
In the second case $M=s_3^2c_1c_2$ cannot be zero
since $c_1, c_2=\pm 1$ and $s_3^2>0$.

Finally, if all three $c_1, c_2, c_3$ are different, then from $U=0$ we
have $s_1s_2=s_1s_3=s_2s_3=0$, so at least two $s_j's$ are zero.
Suppose $s_1=s_2=0$, but again then $M=s_3^2c_1c_2$ cannot be zero.
$\;\;\Box$

\end{document}